# Atomic-scale Visualization of Electronic Fluid Flow

Xiaolong Liu[1§], Yi Xue Chong[1§], Rahul Sharma[1,2] and J.C. Séamus Davis[1,3,4,5]

1. *Department of Physics, Cornell University, Ithaca, NY 14850, USA*
2. *Dept. of Physics, University of Maryland, College Park, MD 20742, USA.*
3. *Department of Physics, University College Cork, Cork T12 R5C, IE*
4. *Max-Planck Institute for Chemical Physics of Solids, D-01187 Dresden, DE*
5. *Clarendon Laboratory, University of Oxford, Oxford, OX1 3PU, UK*
§ *These authors contributed equally to this project.*

ABSTRACT: The most essential characteristic of any fluid is the velocity field $v(\boldsymbol{r})$ and this is particularly true for macroscopic quantum fluids[1]. Although rapid advances[2-7] have occurred in quantum fluid $v(\boldsymbol{r})$ imaging[8], the velocity field of a charged superfluid - a superconductor - has never been visualized. Here we use superconductive–tip scanning tunneling microscopy[9,10,11] to image the electron-pair density $\rho_S(\boldsymbol{r})$ and velocity $v_S(\boldsymbol{r})$ fields of the flowing electron-pair fluid in superconducting NbSe$_2$. Imaging $v_S(\boldsymbol{r})$ surrounding a quantized vortex[12,13] finds speeds reaching 10,000 km/hr. Together with independent imaging of $\rho_S(\boldsymbol{r})$ via Josephson tunneling, we visualize the supercurrent density $j_S(\boldsymbol{r}) \equiv \rho_S(\boldsymbol{r})v_S(\boldsymbol{r})$, which peaks above $3 \times 10^7$ A/cm$^2$. The spatial patterns in electronic fluid flow and magneto-hydrodynamics reveal hexagonal structures co-aligned to the crystal lattice and quasiparticle bound states[14], as long anticipated[15-18]. These novel techniques pave the way for electronic fluid flow visualization in many other quantum fluids.

Visualization of quantum fluid dynamics is now at the research frontier. Molecular tagging velocimetry[2-5] allows visualization of $v(r)$ in superfluid [4]He for studies of obstacle quantum turbulence[2], quantized vortex dynamics[3], and thermal-counterflow quantum turbulence[4]. In superfluid [3]He, effects of the Galilean energy-boosted quasiparticle spectrum of the moving superfluid are used to study quantized vortex rings and superfluid turbulence[6]. In atomic-vapor superfluids such as [23]Na, the phonon Doppler-shift allows superflow $v(r)$ to be imaged[7], while the circulation quanta in turbulent superflow are visualized using Bragg scattering[8]. However, although hydrodynamics of electron fluids have recently become of intense interest[19-23], direct atomic-scale visualization of electron fluid flow remains elusive.

An electron-pair fluid in a macroscopic quantum state with many-body wavefunction $\psi(r) = \sqrt{n_S(r)}e^{i\theta(r)}$ is nominally a superconductor[1]. Here $n_S(r)$ is the number density of condensed electron-pairs with charge –2e at location $r$, $\rho_S(r) \equiv -2en_S(r)$ is the electron-pair density, and $\theta(r)$ the macroscopic quantum phase. If $\theta(r)$ varies spatially, this implies that the superfluid is moving relative to the host crystal with superfluid velocity $v_S(r)$ such that $\hbar\nabla\theta(r) = 2mv_S(r) - 2eA(r)$; $2m$ is the effective mass of an electron-pair and $A(r)$ is the vector potential of magnetic fields $B(r)$ due to supercurrent density $j_S(r)$. The quantum fields $\psi(r)$, $\rho_S(r)$ and associated electronic fluid flow fields $v_S(r)$, $j_S(r)$ are characteristics of the two-particle condensate and not of the Bogoliubov quasiparticle excited states[1]. And, while such quasiparticles have been visualized in a wide variety of single-electron tunneling experiments, few visualizations of $\rho_S(r)$ have been achieved[9,10,11] and none whatsoever of $v_S(r)$ or $j_S(r)$ at atomic-scale.

The magneto-hydrodynamics of superconductive flow has an intricate phenomenology. First, there is the Meissner effect wherein an external magnetic field $B$ is completely excluded from the superconductive bulk except for a layer of thickness the London penetration depth[1] $\lambda$ or within topological defects of the order parameter. Second, the electron-pair density $\rho_S(r)$ is itself influenced by the superfluid velocity $v_S(r)$, in the simplest case as $\rho_S(v_S) \cong \rho_S(0)(1 - v_S^2/v_0^2)$ for samples of thickness $t < \xi$, where $v_0 = \hbar/2m\xi$. Finally, in the reference frame which is stationary with respect to the superfluid, each Bogoliubov quasiparticle state $|\mathbf{k}\rangle$ exhibits the standard energy spectrum $E_{\mathbf{k}} = \pm\sqrt{\varepsilon_{\mathbf{k}}^2 + \Delta_{\mathrm{S}}^2}$ (Fig. 1a, red). Here $\varepsilon_{\mathbf{k}}$ is the normal-state band structure referenced to the Fermi energy and $\Delta_{\mathrm{S}}$ is the electron-pairing order parameter. But in the laboratory/crystal frame where the superfluid has velocity $\mathbf{v}_{\mathrm{S}}(\mathbf{r}) \neq 0$, each quasiparticle $|\mathbf{k}\rangle$ receives a Galilean energy-boost[6,24,25,26] yielding a new quasiparticle spectrum $E'_{\mathbf{k}} = \pm\sqrt{\varepsilon_{\mathbf{k}}^2 + \Delta_{\mathrm{S}}^2} + \hbar\mathbf{k}\cdot\mathbf{v}_{\mathrm{S}}(\mathbf{r})$ (Fig. 1a, blue). Thus, in principle, the flow field of an electronic fluid could be visualized by imaging

$$\mathbf{v}_{\mathrm{S}}(\mathbf{r})\cdot\mathbf{n}_{\mathbf{k}_{\mathrm{F}}} = (E'_{\mathbf{k}_{\mathrm{F}}}(\mathbf{r}) - E_{\mathbf{k}_{\mathrm{F}}}(\mathbf{r}))/\hbar k_{\mathrm{F}} \equiv \delta E_{\mathbf{k}_{\mathrm{F}}}(\mathbf{r})/\hbar k_{\mathrm{F}} \quad (1)$$

where $\mathbf{n}_{\mathbf{k}_{\mathrm{F}}}$ is the unit vector of $\mathbf{k}_{\mathrm{F}}$. Therefore, the electronic fluid flow speed can be determined as

$$v_{\mathrm{S}} = max(\delta E_{\mathbf{k}_{\mathrm{F}}})/\hbar k_{\mathrm{F}} \equiv \delta E_{k_{\mathrm{F}}}/\hbar k_{\mathrm{F}} \quad (2)$$

for simple circular Fermi surfaces.

In practice, however, use of Eqn. 2 presents a number of technical barriers. These include requirements to achieve imaging of the electron-pair density $\rho_{\mathrm{S}}(\mathbf{r})$, and µeV resolution imaging of $\delta E_{\mathbf{k}_{\mathrm{F}}}(\mathbf{r})$. Millikelvin scanning tunneling microscopy (STM) using superconductor-insulator-superconductor (SIS) junctions has recently emerged as a key approach. In Josephson tunneling of electron-pairs between a superconducting tip and sample[9,10,11,27,28], the Josephson critical current is $I_{\mathrm{J}} \propto \sqrt{\rho_{\mathrm{T}}\rho_{\mathrm{S}}}/R_{\mathrm{N}}$ as $T \to 0$. Therefore, scanned Josephson tunneling microscopy (SJTM) imaging of $I_{\mathrm{J}}^2(\mathbf{r})R_{\mathrm{N}}^2(\mathbf{r}) \propto \rho_{\mathrm{S}}(\mathbf{r})$ should allow atomic-scale visualization of the electron-pair density[9,10,11]. Here, $\rho_{\mathrm{T}}$ is the constant electron-pair density in the scan tip, and $R_{\mathrm{N}}$ is the normal-state resistance of the Josephson junction. Most SJTM systems operate at temperatures $kT > E_{\mathrm{J}} \equiv \Phi_0 I_{\mathrm{J}}/2\pi$ ($\Phi_0$ is the magnetic flux quantum) such that electron-pair tunneling exhibits a phase-diffusive steady-state current[29] $I_{\mathrm{P}}(V_{\mathrm{J}}) = \frac{1}{2}I_{\mathrm{J}}^2 Z V_{\mathrm{J}}/(V_{\mathrm{J}}^2 + V_{\mathrm{C}}^2)$ (Fig. 1b). $V_{\mathrm{J}}$ is the phase-diffusive voltage across the junction; $Z$ is its high-frequency impedance, and $V_{\mathrm{C}} = 2eZk_{\mathrm{B}}T/\hbar$. Thus $(dI_{\mathrm{P}}/dV_{\mathrm{J}})|_{V_{\mathrm{J}}=0} \equiv g(0) \propto I_{\mathrm{J}}^2$ (Fig. 1b) so that[9,10,11]

$$\rho_{\mathrm{S}}(\mathbf{r}) \propto g(\mathbf{r},0)R_{\mathrm{N}}^2(\mathbf{r}) \quad (3)$$

(Supplementary Note 1). For quasiparticle tunneling between the same tip and sample, the quasiparticle current as $T \to 0$ is

$$I_Q(V) \propto \int_0^{eV} N_T(\epsilon - eV) N_S(\epsilon) d\epsilon \qquad (4)$$

This yields an SIS tunneling spectrum

$$g(V) \equiv \frac{dI_Q}{dV} \propto \frac{d}{dV} \int_0^{eV} N_T(\epsilon - eV) N_S(\epsilon) d\epsilon \qquad (5)$$

Here the quasiparticle density-of-states $N_{T,S}(\epsilon)$ can be determined from the quasiparticle Green's function

$$G(\mathbf{k}, \epsilon, \Delta_{T,S}) \equiv (\epsilon + i\delta + \varepsilon_{\mathbf{k}})/((\epsilon + i\delta)^2 - \varepsilon_{\mathbf{k}}^2 - \Delta_{T,S}^2) \qquad (6)$$

with scattering rate $\delta/\hbar$. Thus,

$$N_{T,S}(\epsilon) = -\int d\mathbf{k} \frac{2}{\pi} \Im[G(\mathbf{k}, \epsilon, \Delta_{T,S})] \qquad (7)$$

where $\Im$ denotes taking the imaginary part, and tip and sample have superconducting energy gaps $\Delta_T$ and $\Delta_S$ respectively. The SIS spectrum $g(V)$, being the convolution of two superconducting coherence peaks diverging at $V = \pm\Delta_T/e$ and $V = \pm\Delta_S/e$, is then an extremely sharply peaked function at $V = \pm(\Delta_T + \Delta_S)/e$ (bottom spectrum in Fig. 1c), and is dominated by states at $\mathbf{k} = \mathbf{k}_F$ (Fig. 1a). However, an electron-pair fluid flowing through the sample with velocity $\mathbf{v}_S$ modifies $N_S(\epsilon)$ due to the Galilean energy-boost of $\delta E_{\mathbf{k}} = \hbar \mathbf{k} \cdot \mathbf{v}_S$. In this case[30],

$$N_S(\epsilon, \delta E_{\mathbf{k}_F}, \Delta_S) = -\int d\mathbf{k} \frac{2}{\pi} \Im[G(\mathbf{k}, \epsilon - \delta E_{\mathbf{k}}, \Delta_S)] \qquad (8)$$

so that $g(V, \Delta_S, \delta E_{\mathbf{k}_F})$ becomes a quite complex function around $V = \pm(\Delta_T + \Delta_S)/e$. Figure 1c shows simulated $g(V, \Delta_S, \delta E_{\mathbf{k}_F})$ spectra derived from Eqn. 5 for a fixed $\Delta_T = 1$ meV and $\Delta_S = 1.2$ meV and for a variety of superfluid velocities $0 < v_S < 1600$ m/s. It is the ultra-high energy-resolution SIS tunneling (Supplementary Fig. 1) which allows the Galilean energy-boosts $\delta E_{\mathbf{k}_F}$ to become manifest as splitting of the sharp maxima in $g(V, \Delta_S, \delta E_{\mathbf{k}_F})$ at $V = \pm(\Delta_T + \Delta_S)/e$ (arrows in Fig. 1c). Note that Fig. 1c is pedagogical, designed to illustrate the isolated effects of Galilean energy-boosts by keeping $\Delta_S$ constant in all spectra, and that Eqn. 8 yields far more complex $g(V, \Delta_S, \delta E_{\mathbf{k}_F})$ spectra for real materials and variable $\Delta_S$. Nonetheless, due to the different roles played by $\Delta_S(\mathbf{r})$ and $\delta E_{\mathbf{k}_F}$ in Eqn. 8, SIS imaging of $g(\mathbf{r}, V)$ with sufficient spatial and energy resolution near $V = \pm(\Delta_T + \Delta_S)/e$ should allow visualization of both $\Delta_S(\mathbf{r})$ and $\mathbf{v}_S(\mathbf{r})$ (Supplementary Note 2). The overall challenge has been to achieve $g(\mathbf{r},0)$ and $\delta E_{k_F}(\mathbf{r})$ imaging adequate to yield simultaneous, two-dimensional visualization of $\rho_S(\mathbf{r})$, $\mathbf{v}_S(\mathbf{r})$ and thus $\mathbf{j}_S(\mathbf{r})$.

To explore these challenges, we use 2H-NbSe$_2$ whose superconductive critical temperature is $T_C \approx 7.2$ K and anisotropic energy gap is $\Delta_S(\mathbf{k}) = \Delta_0\{0.8 + 0.2\cos[6\arctan(k_y/k_x)]\}$ (Supplementary Note 3). This material has a hexagonal layered structure with Se-Se separation $d$ (Supplementary Fig. 2), and a coexisting charge density wave state with in-plane wavevectors $\mathbf{Q}_i \approx \{(1,0);(1/2,\sqrt{3}/2);(-1/2,\sqrt{3}/2)\}2\pi/3a_0$ where $a_0 = \sqrt{3}d/2$. Figure 2a shows the topographic image of the NbSe$_2$ surface with a magnetic field $B$ = 50 mT applied to generate a very low density of quantized vortices.

A rapidly flowing electron-pair fluid surrounds each vortex core[12,13] within which quasiparticle states become bound[14,17]. But the iconic Abrikosov model[12] cannot predict the effects on this flow field of crystal fields, or of multiple electronic bands and their anisotropic superconducting energy gaps. Thus, the quantum magneto-hydrodynamic phenomenology of a superconductive vortex has long been studied using the Eilenberger equations for the Greens fuctions[15], or the Bogoliubov-de Gennes equations[16,17,18]. For NbSe$_2$ specifically, self-consistent solutions of the Bogoliubov-de Gennes equations predict that the electron-pair-field, fluid velocity and current density outside the core should all be hexagonal and, moreover, aligned to both the crystal axes and the symmetry axes of quasiparticle bound-states[17,18]. Experimentally, it has never been possible to explore any of these electron-pair fluid flow predictions by visualization.

To do so, we prepare Nb STM tips by field emission on a Nb target, establishing atomic resolution and typically $\Delta_T \approx 1$ meV that is unperturbed under small magnetic fields (Supplementary Fig. 3). Then, in the same field of view (FOV) as Fig. 2a, the vortex core is revealed by visualizing the quasiparticle bound states [14] that vanish exponentially with decay constant $\xi$ for r>$\xi$, by measuring $g(\mathbf{r},\Delta_T/e)$ at $T$ = 290 mK (Fig. 2b). Subsequently we set the origin of coordinates $\mathbf{r} = (0,0)$ at the symmetry point of the vortex core. Next, a complete set of SIS quasiparticle spectra $g(\mathbf{r},V)$ spanning the range $-3$ mV $< V <$ 3 mV is measured in the same FOV of Fig. 2a, b at constant $R_N(r) = 1$ MΩ and $T = 290$ mK (Supplementary Note 2). Figure 2C shows the typical SIS spectrum $g(r \to \infty, V)$ with sharp convoluted coherence peaks at $V = \pm(\Delta_T + \Delta_0)/e$, along with that of $g(r \to 0, V)$ showing the convolved spectrum of the Nb-tip coherence peaks at $V = \pm(\Delta_T)/e$. Simultaneously, we use SJTM to measure the electron-pair tunneling $g(r, 0)$ peak, with the results shown in Fig. 2d. The electron-pair tunneling spectrum as $V \to 0$ is shown in Supplementary Figs. 4, 5. Finally, it is by combining these techniques that the magneto-hydrodynamics of the electron-pair fluid surrounding the vortex core becomes manifest.

Figure 2e (top panel) shows the measured evolution of these SIS spectra along a radial trajectory $0 < r < 152$ nm. Here, the colour code represents intensity of $g(r, V)$, the vertical axis is radius $r$ and horizontal axis is $V$. Next, we determine the superconducting order parameter $\Delta_S(r)$ and the Galilean energy boost $\delta E_{k_F}$ by fitting these measured $g(r, V)$ at each $r$, to the model $g(V, \Delta_S, \delta E_{k_F})$ spectra derived from Eqns. 5 and 8. The best fit is identified by finding the least-valued, normalized, root-mean-square deviation ($\sigma_N$; Supplementary Note 4). Figure 2e (bottom panel) then shows the evolution of the best-fit $g(V, \Delta_S, \delta E_{k_F})$ along the same radial trajectory $0 < r < 152$ nm. The correspondence of experimental $g(r, V)$ (top panel) and fitted $g(V, \Delta_S, \delta E_{k_F})$ (bottom panel) is excellent. This is exemplified directly in Fig. 2f (Supplementary Fig. 7) using examples of experimental $g(r, V)$ and their best-fit $g(V, \Delta_S, \delta E_{k_F})$ at the radii indicated by coloured dashed lines in Fig. 2e. The quality of all the fits in Fig. 2e is quantified by their high coefficients of determination $R^2$ (Supplementary Figs. 8). Next, the best-fit $g(V, \Delta_S, \delta E_{k_F})$ at each $r$ yields $\Delta_0(r)$ and $\delta E_{k_F}(r)$, as shown in Fig. 2g,h respectively. The radial dependence of electron-pair density $\bar{\rho}_S(r)$ is then evaluated from an azimuthal average of $\rho_S(r) \propto g(r, 0)$ (Fig. 3a), while the radial dependence of $\overline{\Delta_0^2}(r)$ is determined from an azimuthal average of $\Delta_0(r)$ (Fig. 3a). By fitting $\bar{\rho}_S(r)/\bar{\rho}_S(\infty)$ to $\tanh^2\left(\sqrt{3/8}\frac{r}{\xi}\right)$ (Supplementary Note 5), we obtain an in-plane coherence length $\xi = 11$ nm, as expected. Then, with superconducting coherence length $\xi \equiv \hbar^2 k_F/\pi m \Delta_0$ and $\Delta_0 = 1.24$ meV, the Fermi wavevector is $k_F = 5.7 \times 10^8 \; m^{-1}$. Finally, the image of electron-pair fluid velocity is attained as $v_S(r) \equiv \delta E_{k_F}(r)/\hbar k_F$. From this, the radial dependence of superfluid speed $\bar{v}_S(r)$ and Galilean energy boost $\overline{\delta E_{k_F}}(r)$ from azimuthal averages of $v_S(\mathbf{r})$ and $\delta E_{k_F}(\mathbf{r})$ is shown in Fig. 3b, for 30 nm $\leq r \leq$ 140 nm (beyond the range of influence of quasiparticle states bound in the core; Supplementary Note 4, Supplementary Fig. 9). Measurements of $v_S(\mathbf{r})$ from different vortices under different magnetic fields and using different SJTM tips yield repeatable, quantitatively indistinguishable, results to those presented here (Supplementary Fig. 10).

Quantitative analysis begins with a comparison between $\overline{\Delta_0^2}(\mathbf{r})$ from fitting SIS $g(\mathbf{r}, V)$ spectra, and the independently determined $\bar{\rho}_S(r)$ from SJTM electron-pair tunneling, finding them in good agreement (Fig. 3a). Because in theory $\rho_S(r) \propto \Delta_0^2(r)$, this observation gives strong confidence in the fitting procedures of the measured SIS $g(V)$ spectra. Next, because of the nearly constant $\rho_S(r)$ in the range of 30 nm $\leq r \leq$ 140 nm, we fit $\bar{v}_S(r) \propto K_1\left(\frac{r}{\lambda}\right)$, where $K_1(x)$ is first-order modified Bessel function of the second kind (Fig. 3b and Supplementary Note 6). This yields an in-plane penetration depth of $\lambda = 160$ nm and thus anisotropy parameter $\kappa_\perp = \lambda/\xi = 14.5$ in agreement with previous reports (Supplementary Note 3). Using the London penetration depth $\lambda = \sqrt{m/2\mu_0 n_S(\infty)e^2}$, we estimate that $n_S(\infty) \cong 5.5 \times 10^{26}/m^3$ yielding $\rho_S(\mathbf{r}) = -2en_S(\infty)g(\mathbf{r},0)/g(r \gg \xi, 0)$. The radial dependence of current density $\bar{j}_S(r)$ as determined from the azimuthal average of $j_S(\mathbf{r}) = \rho_S(\mathbf{r}) v_S(\mathbf{r})$ is then shown in Fig. 3c. As $r \to 0$, $\bar{j}_S(r)$ is estimated as $\bar{\rho}_S(r)\bar{v}_S(r)$ by using the extrapolated $\bar{v}_S(r)$ (Fig. 3b). Because in general $\nabla\theta(\mathbf{r}) = (\mathbf{A}(\mathbf{r}) - m\mathbf{v}_S(\mathbf{r})/e)(-2\pi/\Phi_0)$, the quantum phase winding around the vortex core at radius $r$ is

$$\Theta(r) \equiv |\oint \nabla\theta(\mathbf{r}) \cdot d\mathbf{l}| = \frac{2\pi}{\Phi_0}\left[\iint \mathbf{B}d\mathbf{s} - \oint \frac{m\mathbf{v}_S(r)}{e}d\mathbf{l}\right] \qquad (9)$$

The topological constraint $\Theta = 2\pi$ then generates fluxoid quantization

$$\Phi \equiv \iint \mathbf{B}d\mathbf{s} - \oint \frac{m\mathbf{v}_S(r)}{e}d\mathbf{l} = \Phi_0 \qquad (10)$$

From our measured $\bar{v}_S(r)$ and $\bar{\rho}_S(r)$ these two contributing terms yield

$$\Phi_v(r) \equiv -\oint \frac{m\bar{v}_S(r)}{e}d\mathbf{l} \qquad (11a)$$

$$\Phi_j(r) \equiv \iint \mathbf{B}d\mathbf{s} = 2\pi \int_0^r r_0 dr_0 \int_{r_0}^\infty \mu_0 \bar{v}_S(r') \bar{\rho}_S(r') \, dr' \qquad (11b)$$

where Eqn. 11b is derived using Ampere' Law with azimuthal symmetry. The measured $\Phi_v(r)$ and $\Phi_j(r)$ are shown in Fig. 3d as blue and red circles respectively. As anticipated, they evolve in opposite directions such that their sum results in virtually radius-independent value $\Phi(r) = (0.85 \pm 0.1)\Phi_0$ (white circles Fig. 3d). Together with the high SIS spectra fitting quality ($\langle R^2 \rangle = 0.96$, $\sigma_N \approx 5\%$, Supplementary Figs. 8-9), the close matching of $\overline{\Delta}_0^2(r)$ with $\bar{\rho}_S(r)$ (Fig. 3a), and the agreement between mesured $\xi, \lambda$, and $\kappa_\perp$ with literature values (Supplementary Note 3), these observations demonstrate the validity and internal consistency in the techniques used to visualize and quantify $\Delta_S(\boldsymbol{r}), \rho_S(\boldsymbol{r}), \boldsymbol{v}_S(\boldsymbol{r})$ and $\boldsymbol{j}_S(\boldsymbol{r})$.

Directly imaged flow configurations $\boldsymbol{v}_S(\boldsymbol{r})$ and $\boldsymbol{j}_S(\boldsymbol{r})$ for the quantum vortex of NbSe$_2$ are shown in Fig. 4a and b, respectively (3D representations are shown in Fig. 4c,d). While measured electron-pair fluid speeds diverge to $v_S(\boldsymbol{r}) > 2{,}800$ m/s ($> 10{,}000$ km/hr) as $r \to 0$, $j_S(\boldsymbol{r})$ initially rises but is driven to zero by falling $\rho_S(\boldsymbol{r})$ so that peak values reach approximately $3 \times 10^7$ A/cm$^2$. Evaluation of the fluxoid from measured $\boldsymbol{v}_S(\boldsymbol{r})$ and $\boldsymbol{j}_S(\boldsymbol{r})$ yields a value slightly less than $\Phi_0$, perhaps because of deviations of magnetic field direction away from the z-axis due to surface termination of $\boldsymbol{j}_S(\boldsymbol{r})$. More importantly, the electron-pair fluid velocity field $v_S(\boldsymbol{r})$ exhibits a distinct hexagonal symmetry aligned to the crystal axis (Fig. 4a), and that this phenomenon is more pronounced in $j_S(\boldsymbol{r})$ (Fig. 4b, Supplementary Fig. 11). Moreover, the image of fitted $\Delta_S(\boldsymbol{r})$ evidences hexagonal symmetry (Fig. 2g, Supplementary Fig. 12). Hence, surrounding each NbSe$_2$ vortex core the imaging reveals an electron-pair-potential, a velocity field for the electron-pair fluid flow, and a pattern of current density, all of which are hexagonal. They are all co-aligned to both the crystal axes and to the quasiparticle bound states (Fig. 2b, Supplementary Fig. 12), as long anticipated[15,16,17,18].

Overall, by introducing techniques for simultaneous imaging of $\boldsymbol{v}_S(\boldsymbol{r}), \rho_S(\boldsymbol{r})$ and thus $\boldsymbol{j}_S(\boldsymbol{r})$ of a flowing electronic fluid, we visualize the atomic-scale magneto-hydrodynamics surrounding a superconductive quantized vortex core (Fig. 4), finding it in excellent agreement with long-standing theories. Novel research prospects thus revealed include to visualize flowing electronic fluids surrounding vortices in topological and cuprate superconductors, in the surface currents generated by the superconductive Meissner effect, in chiral edge currents of topological superconductors, and in the viscosity-influenced currents of ultra-metals (Supplementary note 7). More generally, visualization is an extremely powerful tool for scientific research, and the capability to visualize flowing electronic fluids introduced here holds equivalent promise.

**Methods**

A custom-built scanned Josephson tunneling microscope (SJTM) with a base temperature of ~290 mK is used to measure high-quality NbSe$_2$ single crystals (HQ Graphene). SPECS Nanonis electronics are used for data acquisition protocols. Crystals are cleaved *in situ* in cryogenic ultrahigh vacuum at ~4.2 K, and immediately inserted into the STM head. To create the superconductive vortices a magnetic field of 50 mT (unless otherwise noted) is applied normal to the crystal surface. The superconducting Nb tips are prepared by field emission of a fine Nb wire on a Nb target. Topographic images, $T(\boldsymbol{r},V)$, are acquired in constant-current mode under a sample bias of *V*. Differential tunneling conductance images are acquired using a lock-in amplifier (Stanford Research SR830) with a bias modulation of 50 μV. Detailed theoretical analysis and modeling procedures are given in Supplementary Notes 1-7.

**Acknowledgements**

The authors acknowledge and thank J.E. Hoffman, H. Suderow and Z. Hadzibabic for very helpful discussions and advice. X.L. acknowledges support from Kavli Institute at Cornell. X.L., Y.X.C., R.S. and J.C.S.D. acknowledge support from the Moore Foundation's EPiQS Initiative through Grant GBMF9457. J.C.S.D. acknowledges support from the Royal Society through Award R64897, from Science Foundation Ireland under Award SFI 17/RP/5445, and from the European Research Council (ERC) under Award DLV-788932.

**Author Contributions**

X.L and Y.X.C. carried out the experiments; X.L., Y.X.C., and R.S. developed and implemented analysis. J.C.S.D. conceived and directed the project. The paper reflects contributions and ideas of all authors.

**Competing Interests**

The authors declare no competing interests.

**Additional Information**

Supplementary information is available in the online version of the paper. Reprints and permissions information is available online at www.nature.com/reprints. Correspondence and requests for materials should be addressed to J.C.S.D.

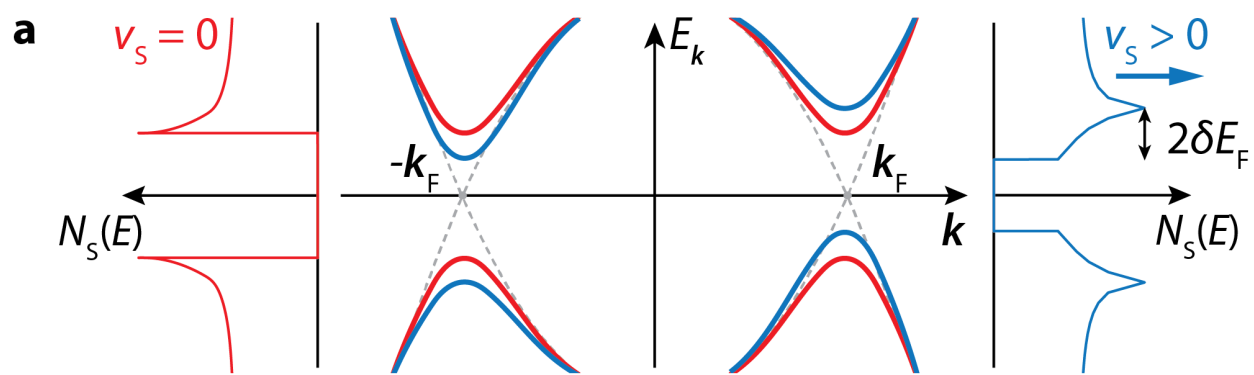
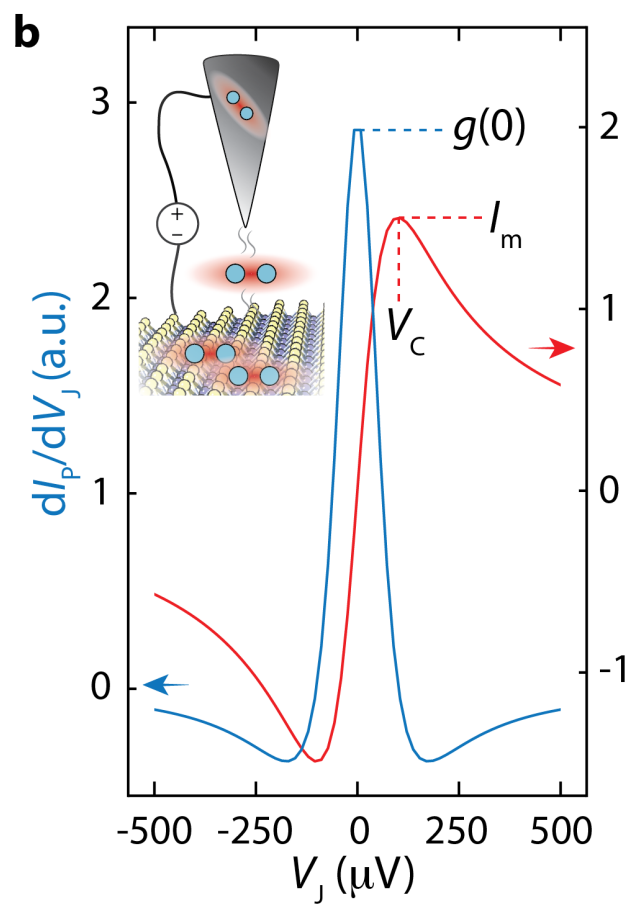
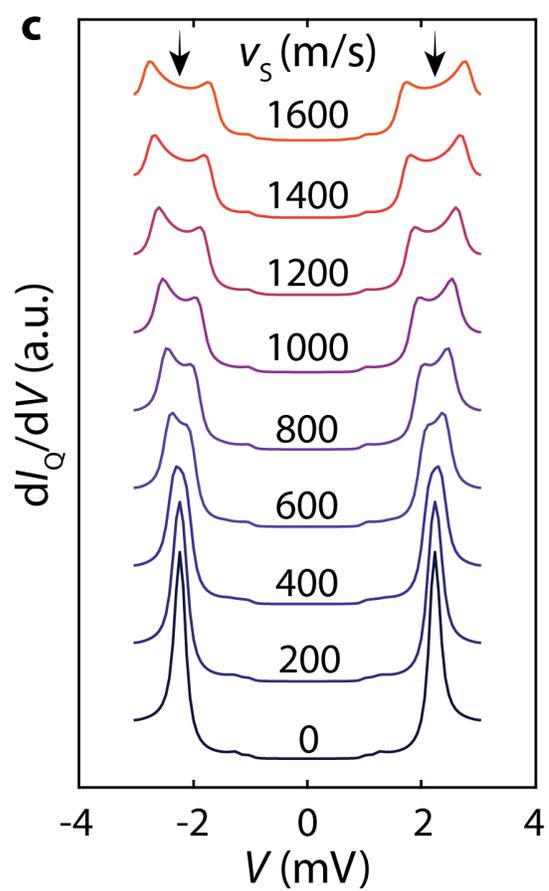

**Figure 1. Quasiparticle and Electron-Pair Tunneling: Model**

**a**, Schematic spectrum $E_{\boldsymbol{k}}$ and density of states $N_S(E)$ of Bogoliubov quasiparticles with electronic fluid velocity $\boldsymbol{v}_S = 0$ (red) and $\boldsymbol{v}_S \neq 0$ (blue) such that $\delta E_{k_F} = 0.4\Delta_S$.

**b**, Phase diffusive Josephson electron-pair spectrum $dI_P/V_J$ (blue) and $I_P(V_J)$ (red). The maximum Josephson current is $I_m$ at voltage $V_J = V_C$. The zero-bias conductance is $g(0)$.

**c**, A series of simulated spectrum $g(V, \Delta_S = 1\,\text{meV}, \delta E_{k_F}) \equiv dI_Q/dV$ in the SIS configuration schematic of Bogoliubov quasiparticle tunneling for electronic fluid flow with different speeds $v_S = \delta E_{k_F}/\hbar k_F$. Comparison between these pedagogical spectra with experimental data should not yield any specific conclusions about $\boldsymbol{v}_S$ in NbSe$_2$, given this simplified pedagogical presentation. Instead, fitting of experimental spectra with realistic modeling of NbSe$_2$ is required (see Supplementary Notes 3, 4).

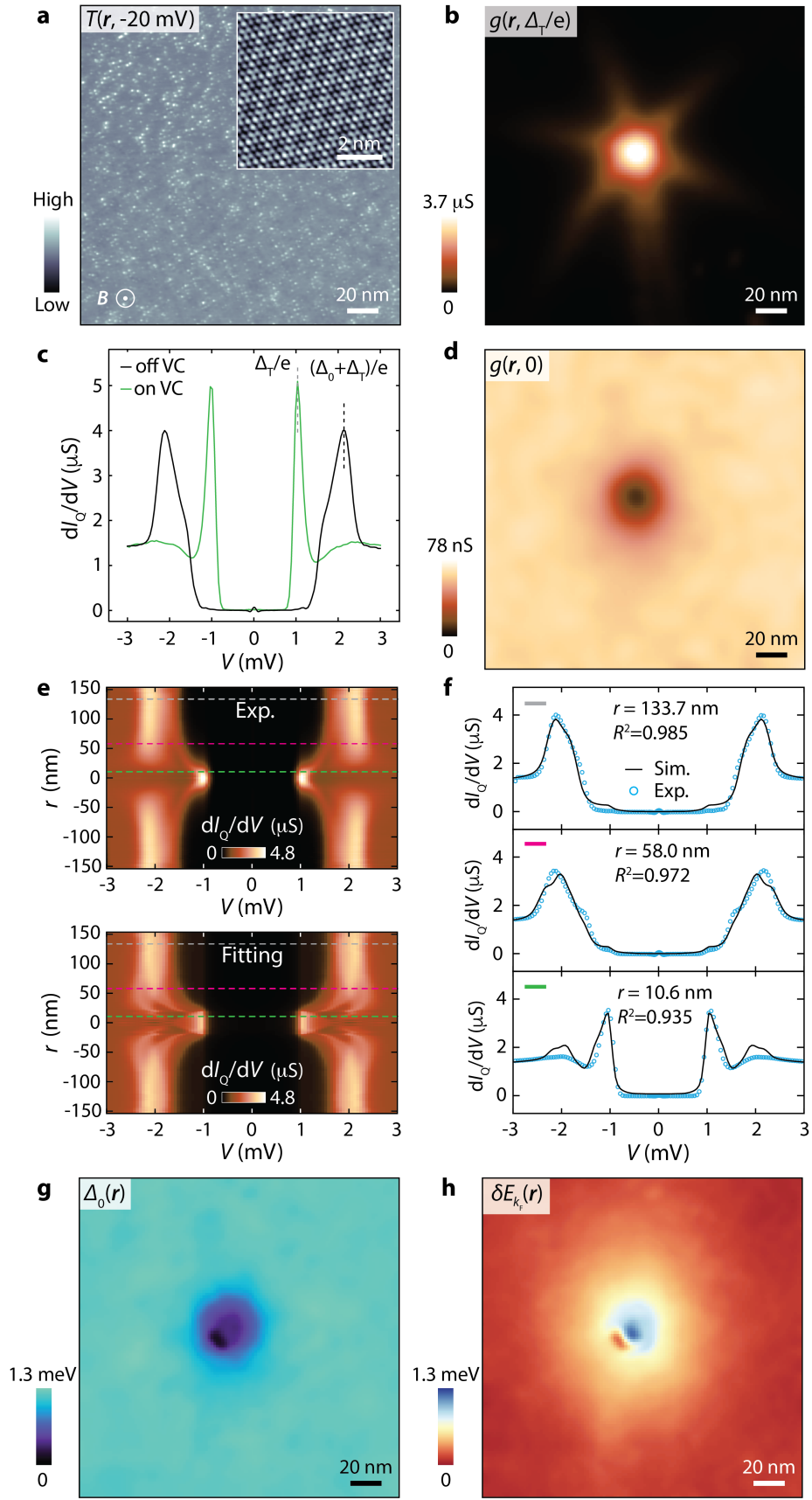

**Figure 2. Quasiparticle and Electron-Pair Tunneling: Experiment**

Images and spectroscopy of a single vortex, all acquired in the same FOV at $T$ = 290 mK and with $B$ = 50 mT applied perpendicular to the surface. Due to SIS tunneling spectroscopy, energy resolution ~20 µeV is achieved throughout these studies.

- **a**, Measured topography $T(r, -20\,\text{mV})$ of studied FOV with inset showing atomically resolved topography under the same tip condition. The white spots are point defects intrinsic to NbSe$_2$.
- **b**, Measured $g(r, \Delta_\text{T}/e)$ revealing the quasiparticle bound states at the vortex core.
- **c**, Measured quasiparticle $dI_\text{Q}/dV$ spectra at $r \to 0$ (green) and $r > 100\,\text{nm}$ (black).
- **d**, Measured $g(r, 0)$ of electron-pair tunneling.
- **e**, Measured $g(r, V) = dI_\text{Q}/dV(r, V)$ spectra for each radius $r$ (top panel). The theoretical $g(V, \Delta_\text{S}, \delta E_{\boldsymbol{k}})$ spectra that best fit the measured $g(r, V)$ at each radius $r$ (bottom panel). Both panels contain the azimuthally averaged results. Multiple Andreev reflections are ruled out as the origins of any of these spectra features (Supplementary Note 4, Supplementary Fig. 6).
- **f**, Examples of fitted $g(V, \Delta_\text{S}, \delta E_{\boldsymbol{k}})$ spectra at different radii indicated by the coloured dashed lines in (e) with coefficients of determination ($R^2$) indicated. The fitting quality only significantly decreases for $r \leq \xi$ close to the vortex core (Supplementary Figs. 8-9).
- **g**, Fitted $\Delta_0(r)$ from the measured $g(r, V)$ spectra.
- **h**, Fitted $\delta E_{k_\text{F}}(r)$ from the measured $g(r, V)$ spectra. The SIS tunneling scheme makes it possible to image Galilean energy-boost spatially because of its greatly enhanced energy resolution[11] compared to single-electron (non-superconductive tip) tunneling at the same temperature. The anomaly near the vortex core is due to a small region of inferior fitting to experimental spectra. To suppress noise, Gaussian-blur by 1.5 pixels is applied to the raw data to generate **b**, **d**, **g**, **h**. The full data acquisition time for the experiment reported here is around 40 hours.

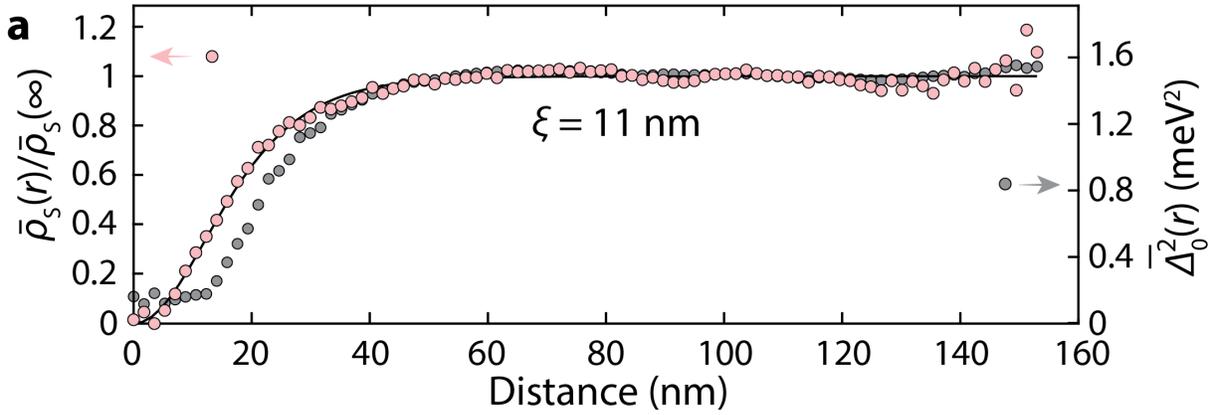
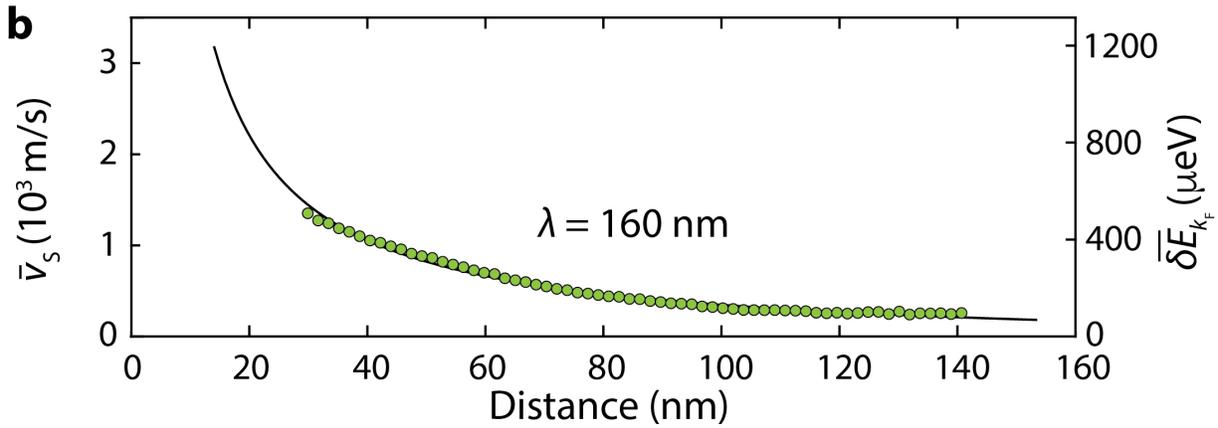
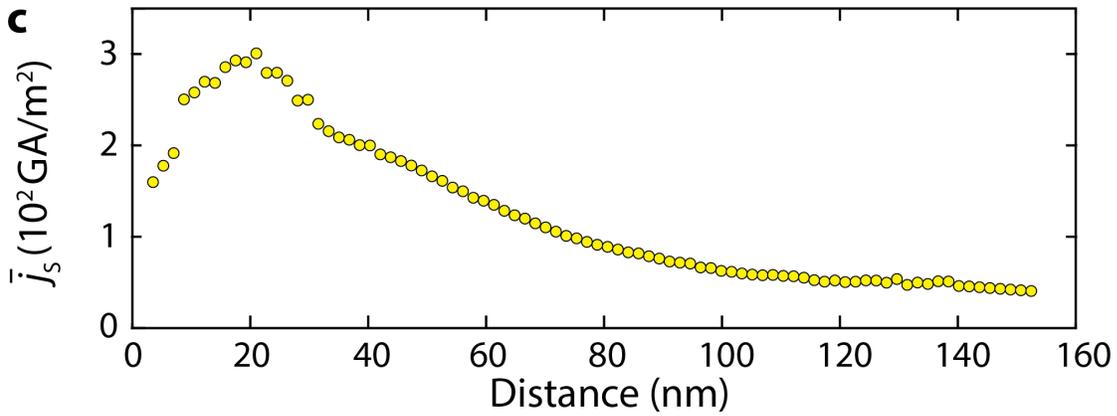
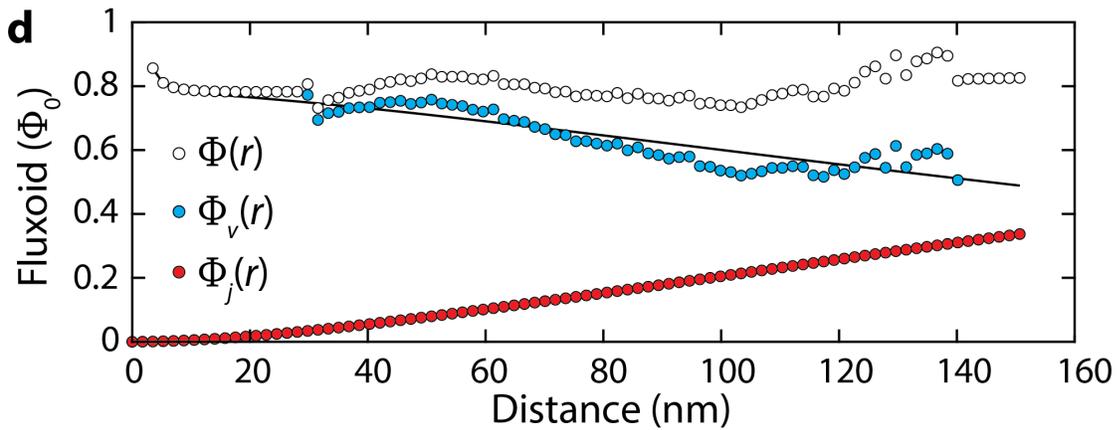

**Figure 3. Radial Dependence of $\rho_S, \Delta_0^2, v_S, j_S,$ and $\Phi$**

**a**, Azimuthal averages of measured $\bar{\rho}_S(r)/\bar{\rho}_S(\infty)$ and $\bar{\Delta}_0^2(r)$. Fitting $\bar{\rho}_S(r)/\bar{\rho}_S(\infty)$ with $\left[\tanh(\sqrt{3/8}\,r/\xi)\right]^2$ as shown results in $\xi = 11$ nm.

**b**, Azimuthal average of measured vortical electronic fluid speed $\bar{v}_S(r)$ in the range of 30 nm $< r <$ 140 nm. Fitting $\bar{v}_S(r)$ with $\bar{v}_S(r) \sim K_1\left(\frac{r}{\lambda}\right)$ results in $\lambda = 160$ nm.

**c**, Azimuthal average of measured vortical current density $\bar{j}_S(r)$. In the range of 30 nm $< r <$ 140 nm, $\bar{j}_S(r) = \overline{\rho_S(r)v_S(r)}$ using measured $\rho_S(r)$ in **a** and $v_S(r)$. For $r <$ 30 nm and $r >$ 140 nm, $\bar{j}_S(r) = \bar{\rho}_S(r)\bar{v}_S(r)$ using measured $\rho_S(r)$ and extrapolated $v_S(r)$ from fitting in **b**.

**d**, Measured $\Phi_v(r)$, $\Phi_j(r)$, and fluxoid $\Phi(r)$. When summed, $\Phi(r) = \Phi_v(r) + \Phi_j(r) \cong (0.85 \pm 0.1)\Phi_0$. The small deviation from $\Phi_0$ is likely due to vortex core bound states (Supplementary Note 4) and modeling $j(r)$ as if around an infinitely long azimuthally symmetric vortex line, while using measured data from the crystal surface where such line terminates.

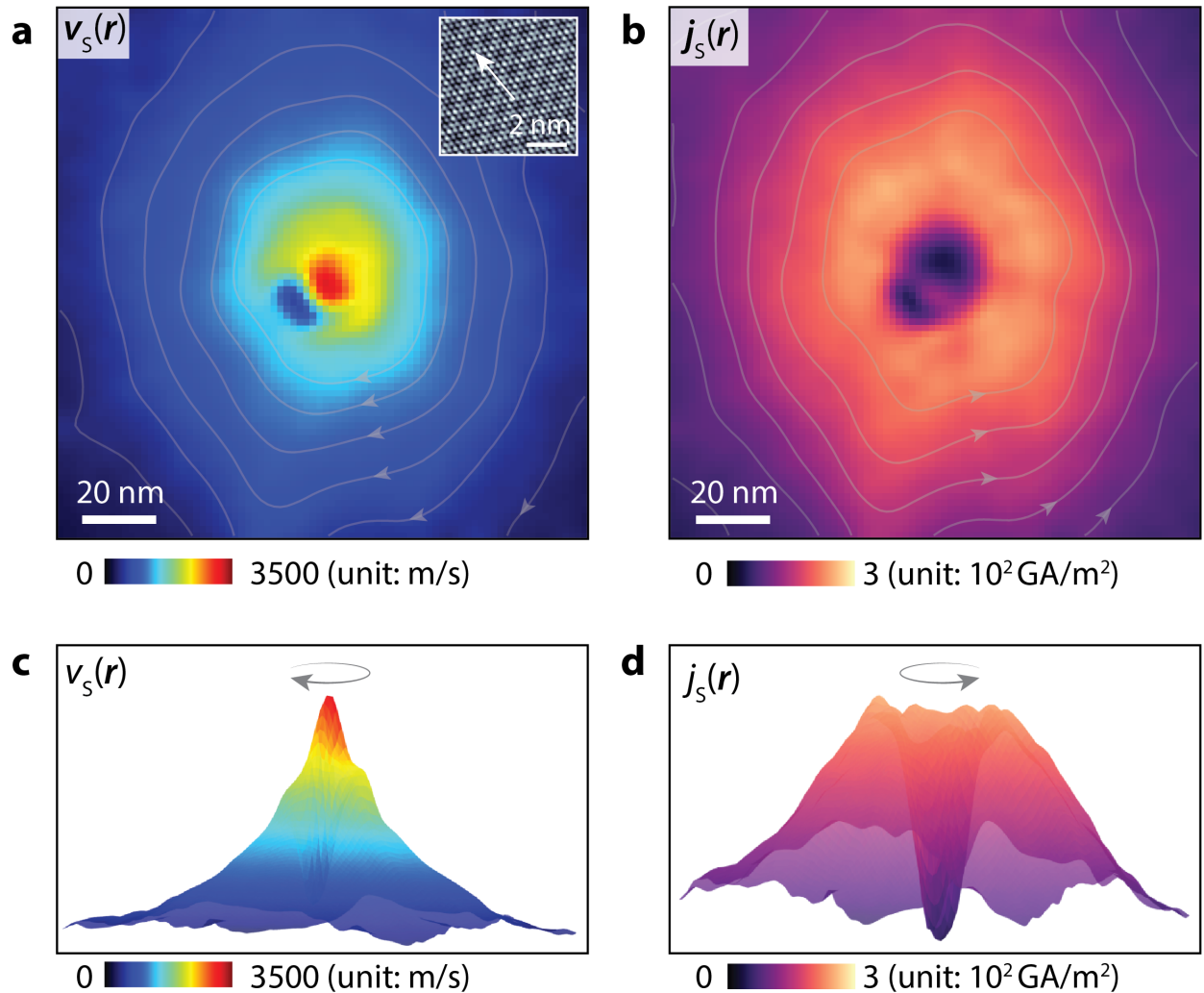

**Figure 4 Visualizing Electronic Fluid Flow**

a, Directly measured $v_S(r)$ without extrapolation with overlaid contour lines from 400 to 1,400 m/s (or 1,440 to 5,040 km/hr) with 200 m/s intervals. The white arrow indicates the lattice direction based on the atomic topography image (inset) taken within the FOV (same as Fig. 2a inset).

b, Directly measured $j_S(r)$ without extrapolation with overlaid contour lines from 90 to 210 GA/m$^2$ with 30 GA/m$^2$ intervals. The FOVs in a and b are 70% of that in Fig. 2a to highlight the vortex. To suppress noise, Gaussian blur by 1.5 pixels is applied to the raw data to generate a, b.

c, Sideview of $v_S(r)$ in a 3D presentation.

d, Sideview of $j_S(r)$ in a 3D presentation.

30 Fulde, P. Gapless superconducting tunneling theory. in *Tunneling phenomena in solids* (ed. E. Burstein, S. Lundqvist), Chapter 29 (Springer, Boston, 1969).

Supplementary Information

# Atomic-scale Visualization of Electronic Fluid Flow

Xiaolong Liu, Yi Xue Chong, Rahul Sharma, and J.C. Séamus Davis

**Supplementary Notes**
**1. Scanned Josephson Tunneling Microscopy**

The macroscopic wavefunctions of a superconducting sample and tip forming a Josephson junction (JJ) can be written as

$$\psi_T = \sqrt{n_T}e^{-i\varphi_T} \tag{S1}$$

$$\psi_S = \sqrt{n_S}e^{-i\varphi_S} \tag{S2}$$

where $n_T$ ($n_S$) and $\varphi_T$ ($\varphi_S$) are the electron-pair number density and the phase of the tip (sample), respectively. The Cooper-pair current ($I_P$) depends on both the phase difference ($\varphi = \varphi_S - \varphi_T$) and the Josephson critical current ($I_J$) as

$$I_P = I_J \sin(\varphi) \tag{S3}$$

while the phase dynamics are related to the junction voltage ($V_J$) as

$$d\varphi/dt = 2eV_J/\hbar \tag{S4}$$

The product of $I_J$ with the normal-state resistance $R_N$ of the JJ is

$$I_J R_N \propto |\psi_T||\psi_S| \propto \sqrt{n_S}\sqrt{n_T} \tag{S5}$$

Therefore, the superfluid density $\rho_S$ in the sample can be measured (if $\rho_T$ is constant) from the relationship

$$\rho_S = 2en_S \propto (I_J R_N)^2 \tag{S6}$$

Visualizing the superfluid density $\rho_S(\mathbf{r})$ of the sample may be achieved by spatial imaging of $I_J(\mathbf{r})$ and $R_N(\mathbf{r})$. When the thermal energy is much smaller than the Josephson energy: $k_B T \ll \frac{\hbar}{2e}I_J$, the junction voltage ($V_J$) remains zero. From the Ambegaokar-Baratoff relation

$$I_J R_N = \frac{\pi \Delta}{2e} \tanh\left(\frac{\Delta}{2k_B T}\right) \tag{S7}$$

where $\Delta$ is the superconducting energy gap (~1 meV), one obtains $I_J \approx 1.57$ nA at $T = 290$ mK with $R_N = 1$ MΩ. Although such an $R_N$ is three orders of magnitude smaller than typical junction resistances for conventional STM operations ($R_N \sim 1$ GΩ) and requires extraordinary tip and temperature stability to achieve imaging, the Josephson energy $\frac{\hbar}{2e}I_J = 3.24$ μeV is still much smaller than the thermal energy $k_B T = 25.0$ μeV so that $\varphi(t)$ fluctuates wildly. In this case, however, the electron-pair tunneling exhibits a phase diffusive steady state current, which reaches maximum ($I_m$) at non-zero junction voltages ($V_C$). Such phase-diffusive Josephson tunneling $I_P(V)$ characteristic are given by[31,32]

$$I_\text{P}(V) = \tfrac{1}{2} I_\text{J}^2 Z\, V_\text{J}/(V_\text{J}^2 + V_\text{c}^2) \tag{S8}$$

where $Z = \hbar V_\text{c}/2ekT^*$ is the impedance relevant to repeated re-trapping of the diffusing phase. Physically, $Z$ is the total electromagnetic impedance of all elements and circuitry adjacent to junction[33], and $T^*$ is an effective temperature; therefore, $Z$ and $V_\text{c}$ are taken to be constant during SJTM imaging. The maximum phase-diffusive electron-pair tunneling current ($I_\text{m}$) is:

$$I_\text{m} = \frac{I_\text{J}^2 Z}{4 V_C} \tag{S9}$$

The first derivative of $I_\text{P}(V_\text{J})$ at zero bias is

$$g(0) \equiv \left.\frac{dI_\text{P}}{dV_\text{J}}\right|_{V_\text{J}=0} = \frac{I_\text{J}^2 Z}{2 V_\text{c}^2} = \frac{2 I_\text{m}}{V_\text{c}} \tag{S10}$$

Therefore, based on equation (S6) and (S10), the sample superfluid density can be measured by using

$$\rho_\text{S}(\boldsymbol{r}) \propto \left(I_\text{J}(\boldsymbol{r}) R_\text{N}(\boldsymbol{r})\right)^2 \propto g(\boldsymbol{r}, 0) R_N^2(\boldsymbol{r}) \tag{S11}$$

**2. Experimental Protocols for Velocimetric Imaging**

A number of key advances towards velocimetric imaging have previously been made, including detection of the effects on $g(V)$ in the presence of unidirectional electrical currents[34,35,36], and of the evolution in amplitude of $g(\pm(\Delta_\text{T} + \Delta_\text{S})/e)$ versus radius from the vortex center [37]. But the outstanding challenge has been to achieve atomic-scale $g(\boldsymbol{r}, 0)$ and $\delta E_{k_\text{F}}(\boldsymbol{r})$ imaging adequate to yield simultaneous, two-dimensional visualization of $\rho_\text{S}(\boldsymbol{r})$, $\boldsymbol{v}_\text{S}(\boldsymbol{r})$ and $\boldsymbol{j}_\text{S}(\boldsymbol{r})$.

In our experiments, a single vortex, e.g., Fig. 2 is first located in a FOV of $205 \times 205$ nm$^2$. Then the tip-sample junction resistance is adjusted to $R_\text{N} \sim 1$ MΩ such that both quasiparticle tunneling at meV energy range and phase-diffusive Josephson tunneling at the μeV energy range, can be detected simultaneously. Then, a full range $dI/dV$ spectrum spanning –3 mV to 3 mV with 165 energy layers is acquired at each of the $117 \times 117$ pixels covering the FOV. The bias voltage step is 8 μV between – 0.2 mV and 0.2 mV and 49 μV in the rest of the voltage range. Therefore, we have $R_\text{N}(\boldsymbol{r})$ being constant in the entire FOV so that $\rho_\text{S}(\boldsymbol{r}) \propto g(\boldsymbol{r}, 0) R_N^2(\boldsymbol{r}) \propto g(\boldsymbol{r}, 0)$.

**3. Realistic Modelling of NbSe$_2$**

To model NbSe$_2$ realistically, we used an anisotropic s-wave order parameter

$$\Delta_\text{S}(\boldsymbol{k}) = \Delta_0 [0.8 + 0.2 \cos(6\phi)] \tag{S12}$$

where $\Delta_0 = 1.24$ meV in zero magnetic field and $\phi = \arctan(k_y/k_x)$. The Fermi surface of bulk NbSe$_2$ in the normal state and absent charge density waves is composed of nearly cylindrical hole pockets at $\Gamma$ with primary contributions from Nb $d_{z^2}$ orbitals, a 3D Se $4p_z$ pocket at $\Gamma$, and nearly cylindrical hole pockets at K that are primarily from Nb $d_{x^2-y^2}$ and $d_{xy}$ orbitals (Supplementary Fig. 2). Based on angle resolved photoemission spectroscopy the effective electron mass $m^* \sim m$ (Ref. [38]) which we use throughout. In NbSe$_2$ there are 5 distinct Fermi surface pockets with complex pocket- and $\boldsymbol{k}$-dependent gap opening due to superconductivity and competition with charge density waves. Here, the superconductivity can be effectively modelled by a single anisotropic gap[39] (Eqn. S12) and an effective in-plane BCS coherence length[40] $\xi \equiv \frac{\hbar v_F}{\pi \Delta_0}$. Using $\xi = 11.2$ nm (Fig. 3a) and $\xi = \frac{\hbar v_F}{\pi \Delta_0}$, we obtain an effective Fermi velocity $v_F \approx 6.6 \times 10^4$ m/s and effective Fermi wavevector $k_F = \frac{m^* v_F}{\hbar} = \frac{0.057}{\text{Å}} = 0.196/d$, where $d = 3.44$ Å. Thus, the normal state band structure of NbSe$_2$ is effectively modelled as $\epsilon_k = 322.6k^2 - 12.4$, where $\epsilon_k$ and $k$ are in the units of meV and $1/d$, respectively. In a magnetic field as small as 50 mT, the Zeeman energy is negligible and thus we have ignored it in the expression of $\epsilon_k$.

The measured $\xi = 11.2$ nm and $v_F \approx 6.6 \times 10^4$ m/s are in good agreement with values in the literature ($\xi = 9.5$ nm (Ref. [41]), 11 nm (Ref. [42]); 13 nm (Ref. [43]); $v_F = 8.5 \times 10^4$ m/s (Ref. [44])). As seen in Fig. 3 and detailed in Supplementary Note 6, the measured in-plane penetration depth $\lambda = 160.1$ nm is consistent with reported values in the literature ($\lambda = 135$ nm (Ref. 40), 160 nm (Ref. 42)). The anisotropy parameter $\kappa_\perp = \frac{\lambda}{\xi} = 14.3$ is also agrees with reported values ($\kappa_\perp = 13.5$ (Ref. [45]), 14 (Ref. [42])).

### 4. Fitting Experimental SIS Spectra to Determine Superfluid Flow

Approaching a vortex core, the Galilean energy-boost $\delta E_{k_F} = \hbar k_F v_S$ modifies the superconducting energy gap continuously, which decreases to zero at $r = 0$. For all parameters in the ranges $0 < \Delta_0 < 1.3$ meV and $0 < v_S < 3500$ m/s ($0 < \delta E_{k_F} < 1.3$ meV), SIS tunneling spectra $g(V, \Delta_S, \delta E_{k_F})$ are then simulated using NbSe$_2$ band/gap structure. To do so, we start by using the parameterized normal state band dispersion (Supplementary Note 3) $\epsilon_k = 322.6k^2 - 12.4$ and anisotropic gap structure (Eqn. S12). Then for a particular set of $\delta E_{k_F}$ and $\Delta_0$, we calculate the quasiparticle Green's function

$$G(\boldsymbol{k}, \epsilon, \Delta_S) = (\epsilon + i\delta + \varepsilon_{\boldsymbol{k}})/((\epsilon + i\delta)^2 - \varepsilon_{\boldsymbol{k}}^2 - \Delta_S^2) \tag{S13}$$

and the corresponding density of states

$$N_S(\epsilon, \delta E_{k_F}, \Delta_S) = -\int d\boldsymbol{k} \frac{2}{\pi} \mathfrak{I}[G(\boldsymbol{k}, \epsilon - \delta E_{\boldsymbol{k}}, \Delta_S)] \tag{S14}$$

, where $\mathfrak{I}$ denotes taking the imaginary part. Since $\mathfrak{I}[G(\boldsymbol{k}, \epsilon - \delta E_{\boldsymbol{k}}, \Delta_S)]$ is dominated by values around $\boldsymbol{k} = \boldsymbol{k}_F$, to increase efficiency, the area of integration over the $\boldsymbol{k}$-plane is defined in polar coordinates, bounded by $k_F - \frac{0.033}{d} < k < k_F + \frac{0.033}{d}$. We initialize a $101 \times 101 \times 100$ 3D grid to store $N_S$ for parameters $0 \leq \Delta_0 \leq 1.3$ meV, $0 \leq \delta E_{k_F} \leq 1.3$ meV, and $-3$ meV $\leq \epsilon \leq 3$ meV, respectively. Convoluting each of these spectra with the tip density of states gives us the $g(V, \Delta_S, \delta E_{k_F})$ spectra for the $\Delta_0$ and $v_S$ phase space we are interested in (Supplementary Fig. 7a).

Supplementary Fig. 7a shows the evolution of the maximum differential conductance of simulated spectra

$$\max\{g(V, \Delta_S, \delta E_{k_F})\} \propto \max\{d[\int_0^{eV} N_T(\epsilon - eV) N_S(\epsilon, \Delta_S, v_S) d\epsilon]/dV\} \quad (S15)$$

with $N_T(\epsilon) = -\int d\mathbf{k} \frac{2}{\pi} \Im[G(\mathbf{k}, \epsilon, \Delta_T)]$ in the ranges $\Delta_0$ and $v_S$ as indicated. Then each spectrum from experiment is compared directly with every simulated spectrum in the full ranges of $\Delta_0$ and $v_S$, and the corresponding $(\Delta_0, v_S)$ values with the least normalized root-mean-square deviation ($\sigma_N$) is assigned to it:

$$\sigma_N(\mathbf{r}) \equiv \sqrt{\sum_{i=1}^{N} [\bar{g}(\mathbf{r}, V_i) - g(V_i, \Delta_S(r), \delta E_{k_F})]^2 / N} \Big/ \{\max[\bar{g}(\mathbf{r}, V)] - \min[\bar{g}(\mathbf{r}, V)]\} \quad (S16)$$

where $N = 57$ is the number of voltage values in a simulated spectrum.

Because each measured spectrum $g(V)$ is a convolution of the Galilean boosted density-of-states in the material $N_S(\epsilon, \delta E_{k_F}, \Delta_S)$ with the known tip spectrum $N_T(\epsilon)$ (Eqn. 5), the effects of electronic fluid flow in $\delta E_{k_F}$ are distributed over the whole structure of each $g(\epsilon, \delta E_{k_F}, \Delta_S)$ or $g(V)$ function beyond the tip energy gap ($\pm \Delta_T$). The Josephson pair-tunneling around zero-energy (Supplementary Figs. 4, 5) is not taken into account in simulated spectra and thus the fitting to experimental spectra excludes the voltage range of $-0.25$ mV to $0.25$ mV.

We also note that the spectra features assigned to the effect of flowing electron-pair fluid cannot be explained by multiple Andreev reflections. This is because multiple Andreev reflection process are manifested as symmetric well-defined in-gap conductance peaks (Supplementary Fig. 6). Such features are drastically different from the evolving SIS coherence peaks shown in Fig. 2 and Supplementary Fig. 7. Furthermore, those multiple Andreev reflection peaks are at energies within the energy gap created by the tip ($|\Delta_T| \sim 1$ meV), whereas the features shown in Fig. 2 and Supplementary Fig. 7 for extracting velocity are beyond $\pm 1$ meV, which is due to the distinct mechanisms of formation. Therefore, one can rule out multiple Andreev reflections as origins of the spectral features.

Examples of fit quality between measured and simulated spectra are show in Fig. 2f and Supplementary Fig. 7e. As seen from Supplementary Fig. 7e, the coherence peaks at $V = \pm(\Delta_0 + \Delta_T)/e$ first decrease in height and increase in width at distances far from the vortex core, which are well captured by the simulated spectra. Because there is no further degree of freedom in simulated spectral height after the initial calibration of simulated spectra to match experimental setup condition by requiring $g(-3 \text{ mV}, \Delta_S, \delta E_{k_F}) \equiv$ constant, the close matching of the conductance values of simulated and experimental spectra shown in Fig. 2f and Supplementary Fig. 7 is not only evidence of excellent fitting, but also a strong indicator of the validity of our model. Further decreasing of the distance leads to peak splitting/broadening and eventually another set of coherence peaks grow at $V = \pm\Delta_T/e$ near the vortex core. The fitting quality can be quantitatively seen in Supplementary Fig. 8 from the spatial distribution and histogram of the coefficient of determination:

$$R^2(\boldsymbol{r}) = 1 - \frac{\sum_{i=1}^{N}\left[g(r,V_i)-g\left(V_i,\Delta_S(r),\delta E_{k_F}\right)\right]^2}{\sum_{i=1}^{N}[g(r,V_i)-\langle g(r)\rangle]^2} \tag{S17}$$

The average $<R^2> = 0.96$ suggests excellent fitting quality throughout. For the virtually all areas outside the vortex core (within which the bound states are confined) $R^2(\boldsymbol{r}) \gtrsim 0.96$.

In Supplementary Fig. 9, the normalized root-mean-square deviation as a function of distance to the vortex core is also given, where $\bar{g}(\boldsymbol{r}, V)$ is azimuthally averaged experimental spectra. The normalized root-mean-square deviation is ~5% in the majority of the range outside the vortex core, in agreement with $R^2$ measurement and again indicative of excellent fitting quality. It only significantly increases below a lower bound of $r \approx 30$ nm likely due to the existence of bound states confined to the vortex core, which has a characteristic radius $r \sim \xi = 11$ nm. Above 140 nm, the number of data points from the map is too small. Therefore, we use measured $v_S(r)$ outside the vortex core in the range of 30 nm < r < 140 nm to extrapolate $v_S(r)$ to a larger range down to $r \to 0$ (Fig. 3b, Supplementary Note 6). For materials with more complex electronic structures, knowledge of the Fermi surface and/or energy-gap structure is required to apply this technique, but as discussed in Supplementary Note 7, this poses no limitations on the applicability of the technique.

To further illustrate the detectable effect of electron-pair fluid flow and demonstrate the goodness of fitting, we fitted experimental $g(V)$ spectra by forcing $v_S = 0$. The coefficients of determination $R^2$ using $v_S = 0$ is far smaller than that using $v_S \neq 0$ as shown in Supplementary Fig. 7b. As seen in Supplementary Fig. 7c,d, the fitted spectra with $v_S = 0$ far deviate from experimental $g(V)$ spectra. This is also seen in Supplementary Fig. 7e, where comparisons of a series of experimental $g(V)$ spectra at different radii with fitted spectra using $v_S \neq 0$ and $v_S = 0$ are given. In the inset, the square of error $[g(\epsilon, v_S = 0, \Delta_S) - g(V)]^2$ of fitting with $v_S \neq 0$ and $v_S = 0$ further demonstrate this point.

Albeit the overall good fitting quality, imperfections in the details of the fitted spectra are visible in Fig. 2f and Supplementary Fig. 7. For example, the lowest fitting quality is seen around the vortex core (e.g., Fig. 2f, $r$ = 10.6 nm) mainly due to the presence of vortex-core bound states. The use of more sophisticated approaches (e.g., solutions to self-consistent Bogoliubov–de Gennes equations or Usadel's equations) would likely further improve the fitting quality but require higher computational power. The straightforward approach taken here, however, is shown to be sufficient for reliable extraction of the electron-pair fluid velocity as evidenced by the internal consistencies and cross-checks (Fig. 3, Supplementary Note 3).

## 5. Ginzburg Landau Model for Superconductive Order-Parameter of Vortex

For a type-II superconductor, the Ginzburg-Landau (GL) free energy (Helmholtz free energy) can be written as

$$F = F_\text{n} + a|\psi|^2 + \frac{b}{2}|\psi|^4 + \frac{1}{4m}|(i\hbar\boldsymbol{\nabla} - 2e\boldsymbol{A})\psi|^2 + \frac{\mu|H|^2}{2} \tag{S18}$$

, where $F_\text{n}$ is the normal state free energy, 2m is the effective mass of a Cooper pair, $\psi$ is the order parameter, $\boldsymbol{A}$ is vector field, and $\boldsymbol{H}$ is the magnetic field inside the superconductor ($\boldsymbol{\nabla}\times\boldsymbol{A} = \mu\boldsymbol{H}$). By minimizing the Gibbs free energy $G = F - \mu\boldsymbol{H}\cdot\boldsymbol{H}_0$ in an external field $\boldsymbol{H}_0$, one obtains the two GL equations:

$$a\psi + b|\psi|^2\psi + \frac{1}{4m}(i\hbar\boldsymbol{\nabla} - 2e\boldsymbol{A})^2\psi = 0 \tag{S19}$$

$$\boldsymbol{j} = \frac{\boldsymbol{\nabla}\times(\boldsymbol{\nabla}\times\boldsymbol{A})}{\mu} = \frac{ie\hbar}{2m}(\psi^*\boldsymbol{\nabla}\psi - \psi\boldsymbol{\nabla}\psi^*) - \frac{2e^2}{m}|\psi|^2\boldsymbol{A} \tag{S20}$$

For a single vortex one often assumes $\psi(\boldsymbol{r}) = f(r)e^{i\theta}$ and symmetric gauge: $\boldsymbol{A} = A(r)\hat{\boldsymbol{\theta}}$. For $r \gg \xi, \psi(r) \to$ constant. With $r < \xi$ near the vortex core, $\psi(r)$ can be approximated by a sum of polynomials. By solving the GL equations, one obtains

$$\psi(r) \propto \left[r - \frac{r^3}{8\xi^2} + O(r^5)\right]e^{i\theta} \tag{S21}$$

for $r < \xi$. Because $\tanh(x) = x - \frac{x^3}{3} + O(r^5)$ as $x \to 0$ and $\tanh(x) \to 1$ as $x \to \infty$, $\psi(r)$ can be approximated by $\psi(r) \sim \tanh\left(\sqrt{\frac{3}{8}}\frac{r}{\xi}\right)$ for the entire range $0 < r < \infty$. Therefore,

$$\rho_\text{S}(r) = \rho_\text{S}(\infty)\left[\tanh\left(\sqrt{\frac{3}{8}}\frac{r}{\xi}\right)\right]^2 \tag{S22}$$

This is a classic approximation but one that ignores the actual band structure and the solution of the microscopic gap equations, for each real material.

## 6. Superfluid Velocity Beyond the Core

In the limit where $n_\text{S}(\boldsymbol{r})$ is spatially homogeneous, the solution to London's Equations yields

$$\boldsymbol{\nabla}^2\boldsymbol{B} = \frac{2\mu_0 n_\text{S} e^2}{m}\boldsymbol{B} = \frac{\boldsymbol{B}}{\lambda^2} \tag{S23}$$

with an exact solution in cylindrical coordinates

$$\boldsymbol{B}(\boldsymbol{r}) = \frac{\Phi_0}{2\pi\lambda^2} K_0(r/\lambda)\hat{\boldsymbol{z}} \tag{S24}$$

where $K_0(x)$ is the zeroth-order modified Bessel function of the second kind and $\lambda = \sqrt{m/2\mu_0 n_S e^2}$ is the London's penetration depth. In such a model, the superfluid velocity is approximated by

$$\boldsymbol{v}_S(r) \equiv -\frac{\boldsymbol{j}_S(r)}{2en_S} = -\frac{1}{2e\mu_0 n_S} \nabla \times \boldsymbol{B} = \frac{\hbar}{2m}\frac{d}{dr}\left[K_0\left(\frac{r}{\lambda}\right)\right]\hat{\boldsymbol{\varphi}} = -\frac{\hbar}{2m\lambda} K_1\left(\frac{r}{\lambda}\right)\hat{\boldsymbol{\varphi}} \tag{S25}$$

While this approximation is not valid unless $r > \xi$, and moreover it does not involve the full details of the electronic structure of a real material, it does provide a useful hypothetical form $\sim K_1\left(\frac{r}{\lambda}\right)$ to fit and extrapolate experimentally determined $v_S(r)$. As seen in Fig. 3b, the fitted penetration depth is $\lambda \cong 160$ nm. As seen from Fig. 3b and Fig. 4a, both the extrapolated and experimentally measured $v_S$ approaches 3,000 m/s near the vortex core. It is close to the critical velocity $v_C \equiv \frac{\Delta_0}{\hbar k_F} \approx 3,300$ m/s, which is as expected.

## 7. Potential Applications of Electronic Fluid Visualization

As demonstrated in Fig. 3, the minimum speed measured in this study is $\sim 100$ m/s. Since we have demonstrated energy resolution below 20 μeV (ref. [46]) using the same superconducting tip technique, a minimum detectable speed of ~50 m/s is theoretically achievable for NbSe$_2$ and far smaller for materials with larger Fermi momenta. The application of this technique to other materials will be straightforward, based on Eqn. 1 ($\boldsymbol{v}_S(\boldsymbol{r}) \cdot \boldsymbol{n}_{\boldsymbol{k}_F} = (E'_{\boldsymbol{k}_F}(\boldsymbol{r}) - E_{\boldsymbol{k}_F}(\boldsymbol{r}))/\hbar k_F)$, where $\delta E_{\boldsymbol{k}_F} = E'_{\boldsymbol{k}_F}(\boldsymbol{r}) - E_{\boldsymbol{k}_F}(\boldsymbol{r})$ and $\Delta_S$ are the only two fitting parameters to be determined from experimental $g(V)$ spectra. Although prior knowledge of the Fermi surface is required, determination and parameterization of electronic band structures poses no general limits as they can be routinely achieved using angle-resolved photoemission spectroscopy or density functional theories. Therefore, our electronic fluid flow technique can find wide applications.

Research opportunities may include: the effect of local heterogeneities (e.g. crystal defects, boundaries, nanostructures, electronic circuit elements) on the electron-fluid flow field can now be visualized; the electron-pair fluid flow surrounding vortices in exotic superconductors, e.g., cuprates and chiral superconductors, can be visualized where the flow patterns should differ significantly from those of conventional vortices but have never been observed; the Meissner effect generates surface currents to exclude applied magnetic fields. In type-I superconductors with high $H_{C1}$, surface current density from the Meissner effect is limited by the critical current density ($J_C$), which is $J_C \sim 1.2 \times 10^{12}$ A/m$^2$ for lead[47] so that the the electron-pair fluid speed in London currents reaches $v_S \sim 500$ m/s; in topological superconductors where time reversal symmetry is broken, e.g. chiral $p$-wave, there can be a spontaneous edge current[48,49] where the electron-pair fluid speed $v_S > 150$ m/s; in ultra-metals such as PdCoO$_2$ (ref. [50]), MoP (ref. [51]), or WP$_2$ (ref. [52]), the electrical conductivity at low temperature approaches $\sigma \sim 10^{11}$ S/m (refs. [53], 51, 52) and they have carrier concentrations of $n \sim 10^{27} - 10^{28}/m^3$ so that a moderate electric field of 1 V/m creates electron fluid speeds $v = \frac{\sigma E}{ne}$ up to 600 m/s.

## Supplementary Figures

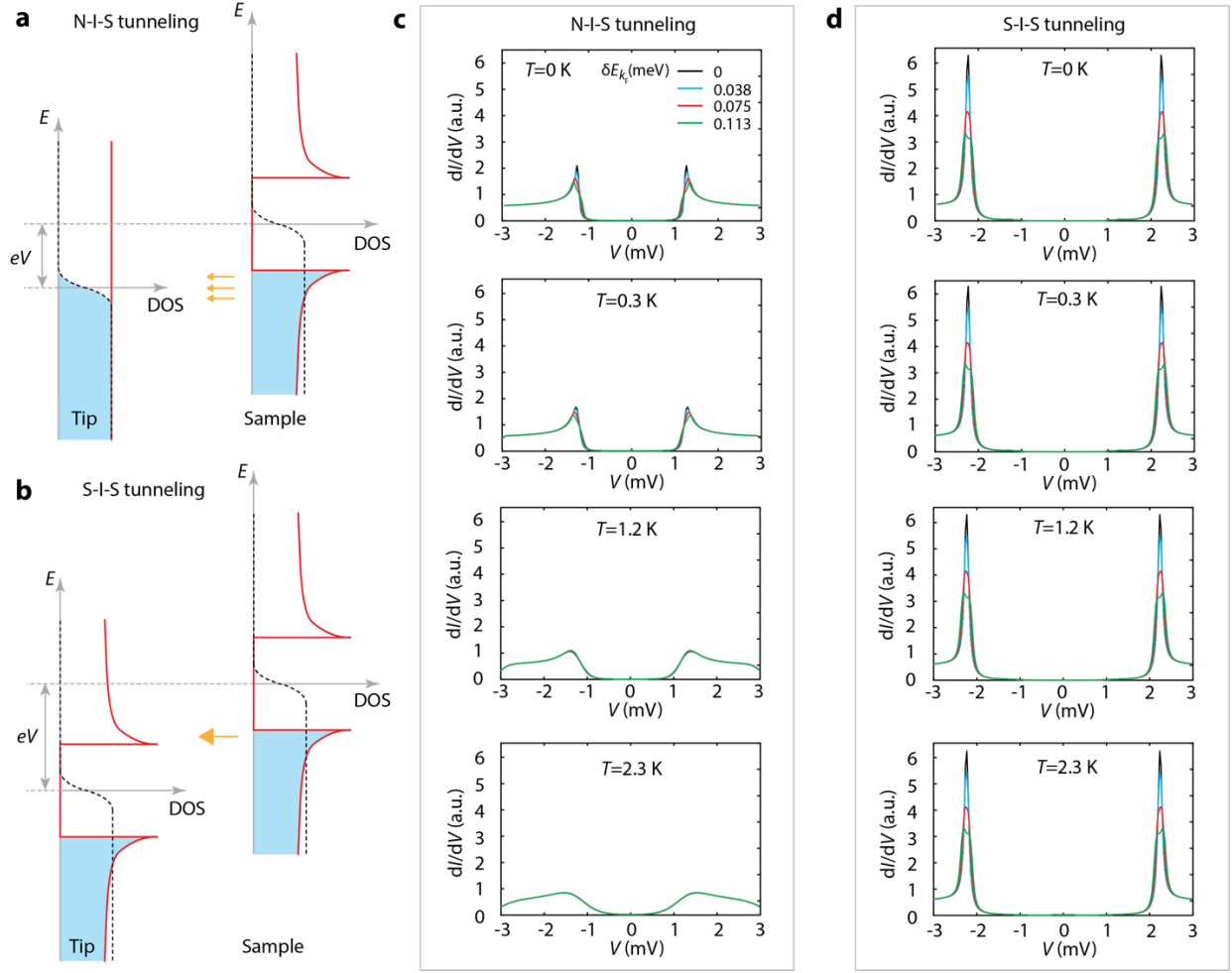

**Supplementary Figure 1. a**, Schematics of band alignments in normal-insulating-superconducting (N-I-S) tunneling with a normal tip and **b**, in superconducting-insulating-superconducting (S-I-S) tunneling with a superconducting tip. The black dashed lines represent Fermi-Dirac distribution at finite temperature. Filled states are colored blue. Quasiparticle tunneling is shown by the yellow arrows. **c**. Simulated differential conductance spectra in the N-I-S and **d**, S-I-S configurations, at different temperature and with different Galilean energy boosts. While thermal broadening at finite temperatures smears the DOS distribution in the normal metal, the presence of a superconductivity gap exceeding thermal energy leads to virtually unaffected tunneling process compared to 0 K in the SIS configuration. Although small Galilean energy boosts can in principle be obtained from N-I-S tunneling spectra at 0 K, as temperature rises to 0.3 K and above, the N-I-S tunneling spectra with different Galilean energy boosts can hardly be distinguished from each other. In contrast, dramatic changes in $g(V)$ are found in simulations within the SIS configuration, providing a clear signature for the determination of Galilean energy boosts. Note that to best compare the capability of N-I-S and S-I-S tunneling, small values of $\delta E_{k_F}$ are used in (c,d),

where the most obvious change of the S-I-S tunneling spectra is the decrease of coherence peak height with increasing $\delta E_{k_\text{F}}$.

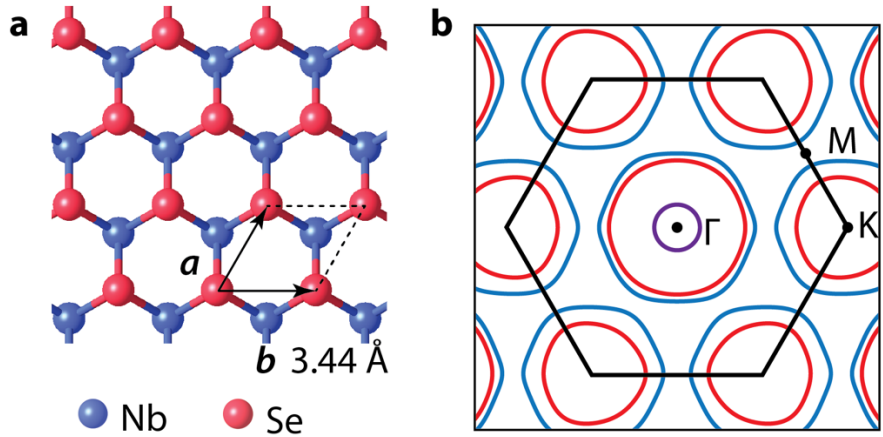

**Supplementary Figure 2. a**, Lattice structure of NbSe$_2$. The lattice constants are $|a| = |b| = d$ = 0.344 nm. **b**, Fermi surface of NbSe$_2$, which is dominated by double-walled nearly cylindrical hole pockets at $\Gamma$ with primary contributions from Nb $d_{z^2}$ orbitals, a 3D Se $4p_z$ pocket at $\Gamma$, and hole pockets at K that are primarily from Nb $d_{x^2-y^2}$ and $d_{xy}$ orbitals.

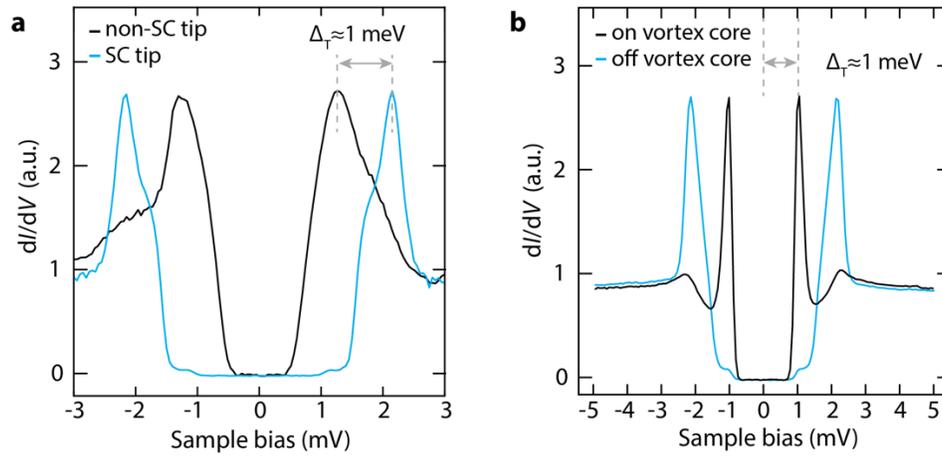

**Supplementary Figure 3. a.** d$I$/d$V$ – $V$ spectra taken on NbSe$_2$ with a superconducting (SC) Nb tip (blue) and a non-SC tip (black) at 0.3K in zero external magnetic field. The tip gap is obtained to be ~ 1 meV. **b**. d$I$/d$V$ – $V$ spectra taken at the vortex core center (black) and far from the vortex core (blue) of NbSe$_2$ under 100 mT magnetic field at 0.3 K. It can be seen that the tip gap remains to be 1 meV. This suggests the tip superconductivity is minimally affected by the small magnetic field from the vortices.

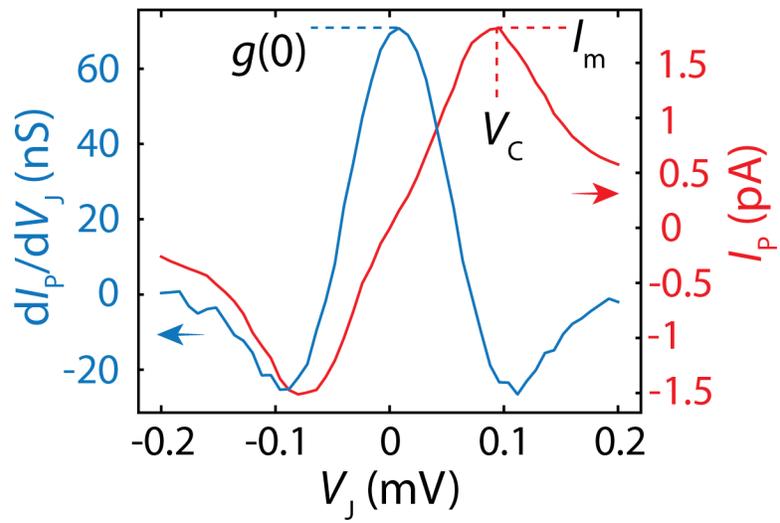

**Supplementary Figure 4**. Measured electron-pair tunneling $dI_P/dV_J$ (blue, using a lock-in amplifier) and $I_P(V_J)$ (red) at $r > 100$ nm.

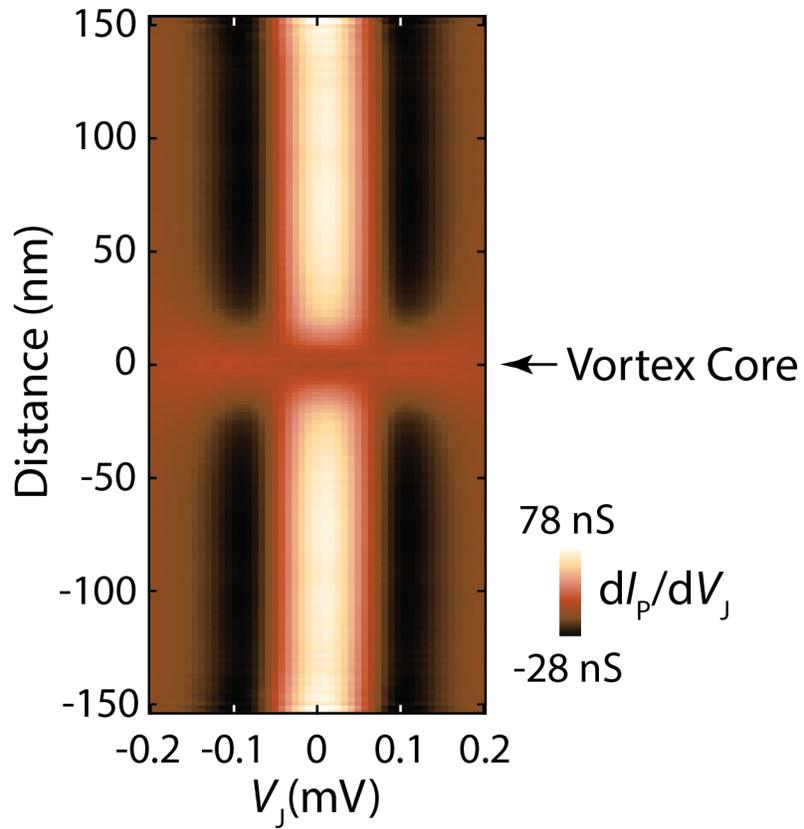

**Supplementary Figure 5.** Azimuthally averaged phase-diffusive Josephson tunneling conductance across the vortex in Fig. 2. The decrease of $g(0)$ near the normal vortex core ($r = 0$) is clearly seen.

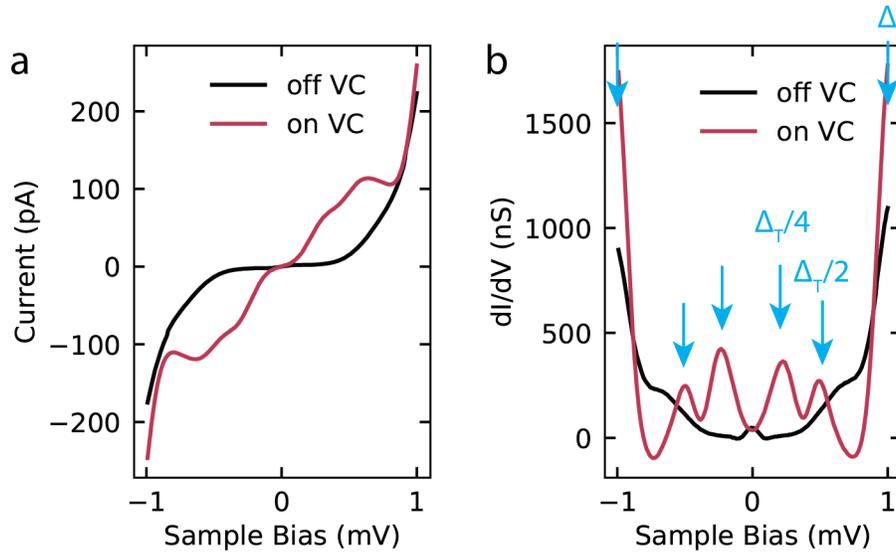

**Supplementary Figure 6. a**, *I-V* and **b**, d*I*/d*V-V* spectra taken with a Nb tip on and off a vortex core of NbSe$_2$. The appearance of multiple Andreev reflection peaks is indicated by the blue arrows.

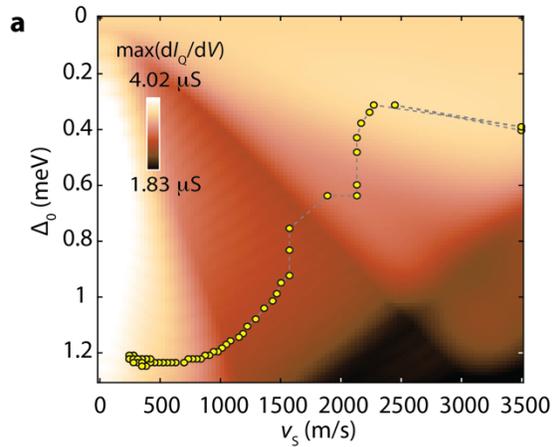
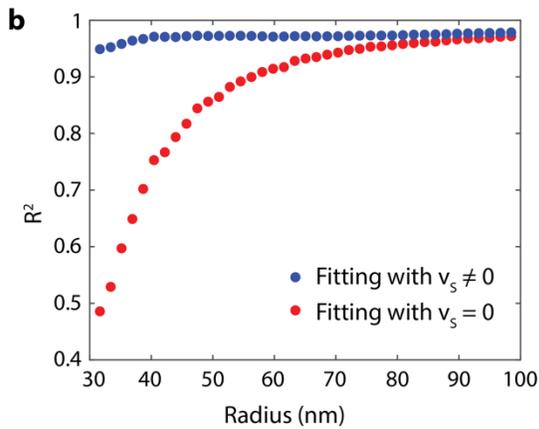
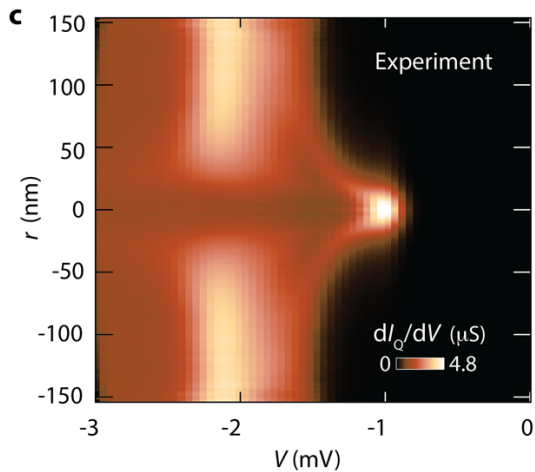
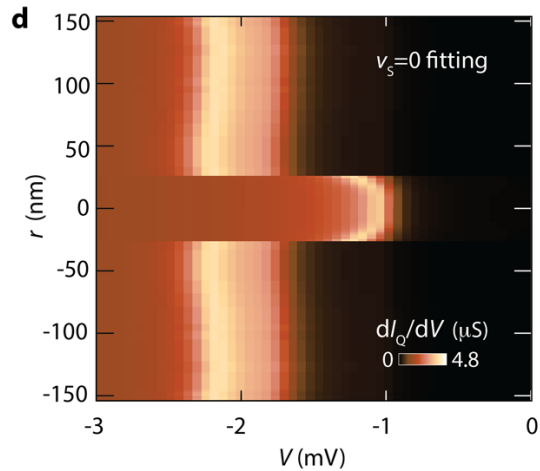
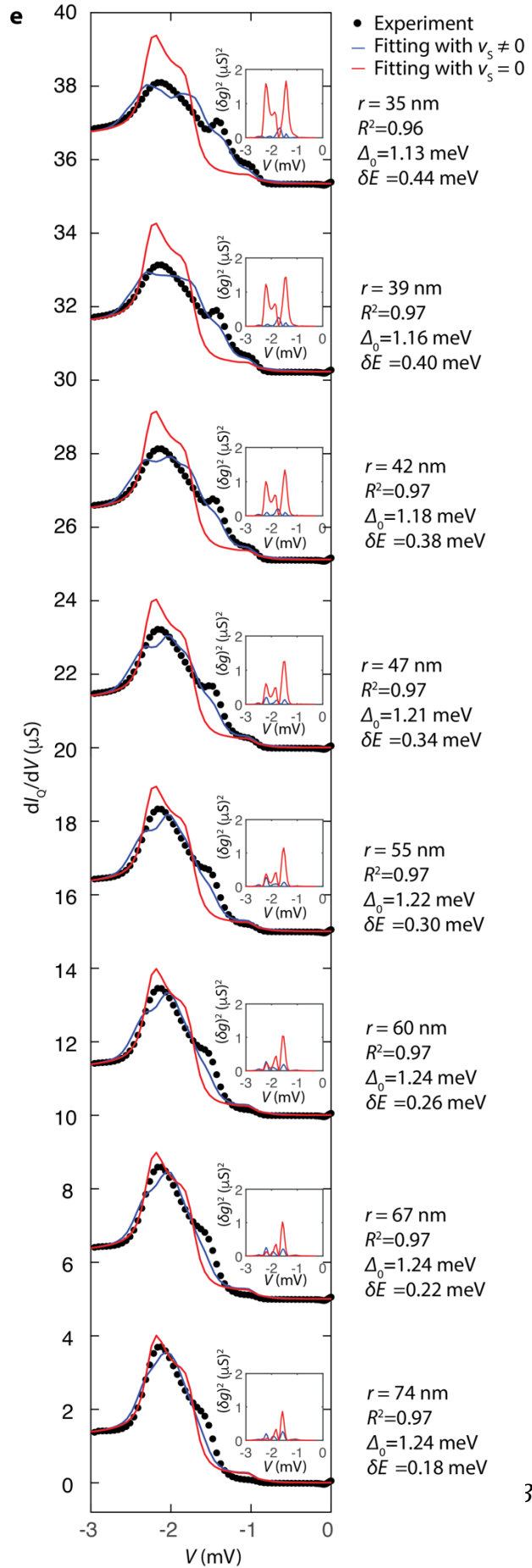

**Supplementary Figure 7. a**, The $\Delta_0 - v_S$ phase space and the trajectory formed by fitting individual azimuthally averaged SIS spectra from experiment. **b**, Comparison of $R^2$ for fitting $g(\epsilon, 0, \Delta_S)$, i.e. with $v_S = 0$ and best fitting of $g(\epsilon, \delta E_{k_F}, \Delta_S)$ with $v_S$ as a free parameter. **c-d.** Color-coded (c) measured $g(r, V)$ spectra for each radius $r$ and fitted spectra using (d) $v_S = 0$. **e.** Series of experimental spectra at different radii with fitted spectra using $v_S = 0$ and $v_S \neq 0$. The inset shows the square of the error in both cases.

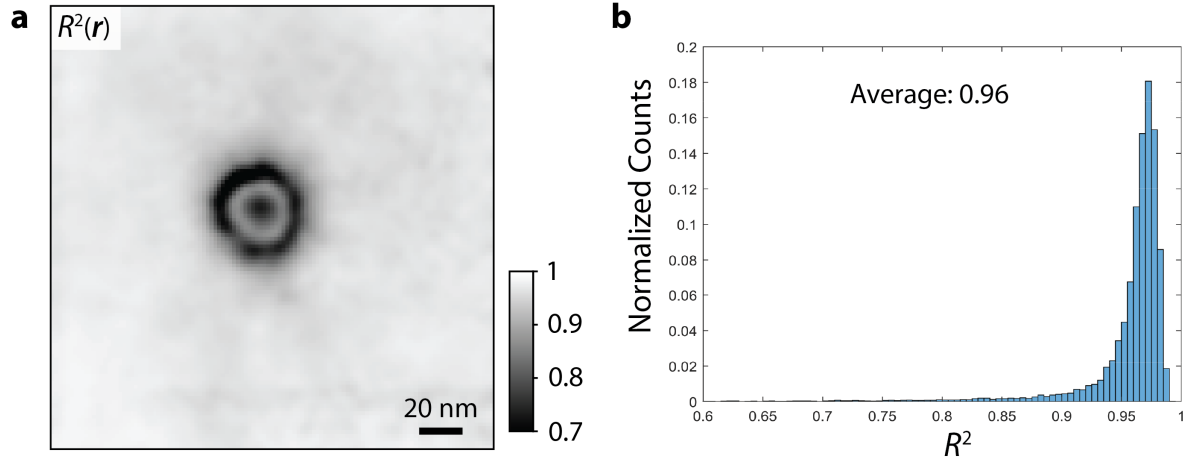

**Supplementary Figure 8. a**, Spatial distribution and **b**, histogram of the coefficient of determination $R^2(r) = 1 - \frac{\sum_{i=1}^{N}\left[g(r,V_i) - g\left(V_i, \Delta_S(r), \delta E_{k_F}\right)\right]^2}{\sum_{i=1}^{N}[g(r,V_i) - \langle g(r)\rangle]^2}$ from the SIS spectra fitting with an average value of $<R^2> = 0.96$. For the majority of the FOV outside the vortex core where bound states are confined, $R^2(r) \gtrsim 0.96$.

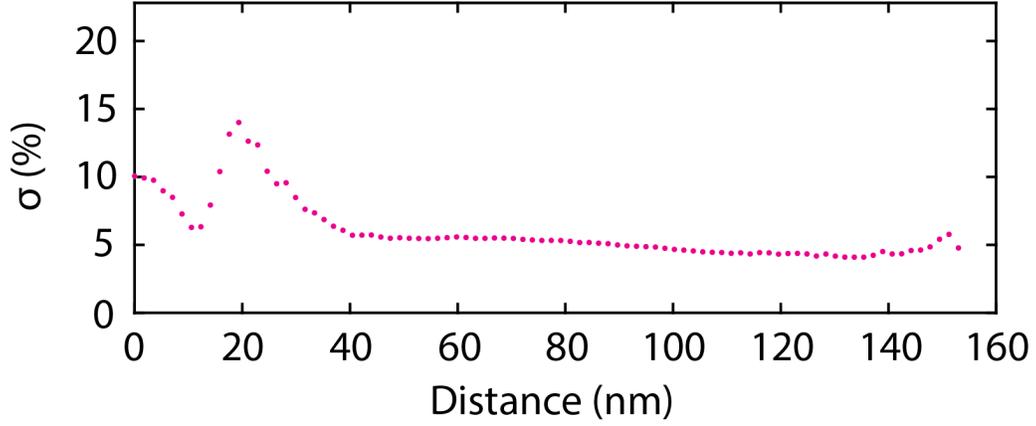

**Supplementary Figure 9.** The normalized root-mean-square deviation

$$\sigma_N(r) \equiv \sqrt{\sum_{i=1}^{N}[\bar{g}(r,V_i) - g(V_i, \Delta_S(r), \delta E_{k_F})]^2 / N} \Big/ \{\max[\bar{g}(r,V)] - \min[\bar{g}(r,V)]\}$$

as a function of distance to the vortex core. Here, $\bar{g}(r,V)$ is azimuthally averaged experimental spectra and $N = 57$ is the number of voltage values in a simulated spectrum. We choose a lower bound $r \approx 30$ nm, below which the normalized root-mean-square deviation increases significantly likely due to the vortex core bound states. Above 140 nm, the number of data points from the map for averaging is too small such that we choose $r \approx$ 140 nm as the upper bound. It can be seen that in the majority of the range outside the vortex core, the deviation is ~5%.

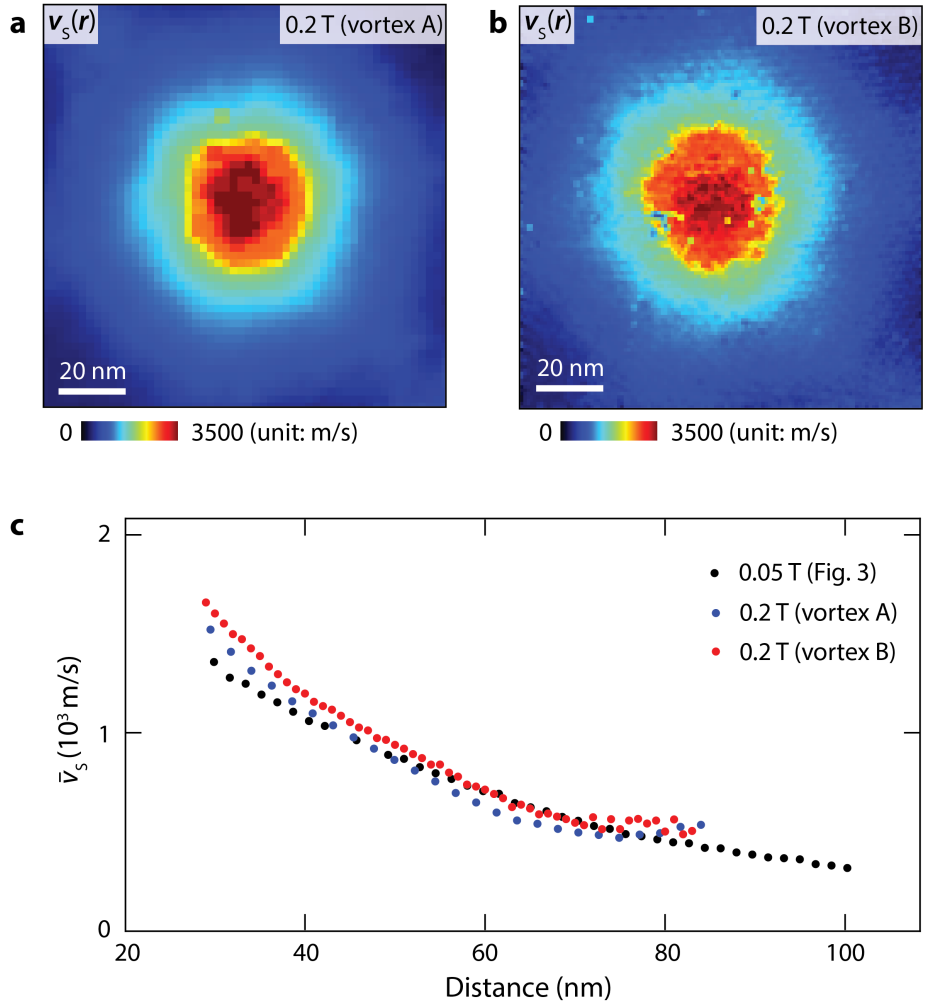

**Supplementary Figure 10. a,b,** Velocity map extracted from another two different NbSe$_2$ vortices formed in a magnetic field of 0.2 T (as opposed to the 0.05 T in Fig. 4). **c,** Comparison of azimuthally averaged velocities as a function of radius showing quantitatively similar results. The close matching of velocities from vortices formed at 0.05 T and 0.2 T suggests the velocity is insensitive to small external magnetic field. The close matching of velocities from two different vortices formed at the same 0.2 T magnetic field suggests the vortices are equivalent. Both of these two observations are consistent with theory and well expected. Finally, while the macroscopic Nb wire remains the same in our SJTM, the microtip formed at the end of the Nb wire is constantly conditioned and sharpened for new experiments. Therefore, the data represent consistent results from different SJTM microtips.

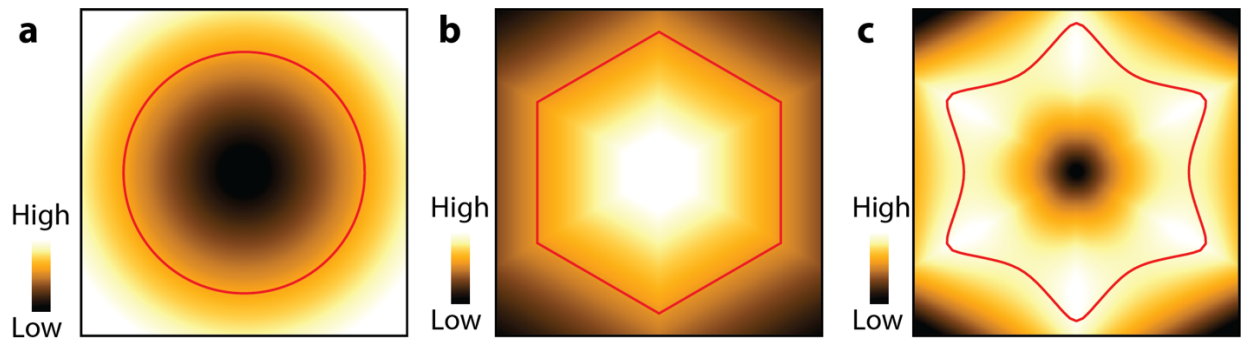

**Supplementary Figure 11.** Schematic illustration of how **a**, a rotationally symmetric $\rho_S(r)$ and **b**, $v_S(r)$ with hexagonal contours produce **c**, $j_S(r) = \rho_S(r)v_S(r)$ with warped contours (i.e., higher anisotropy).

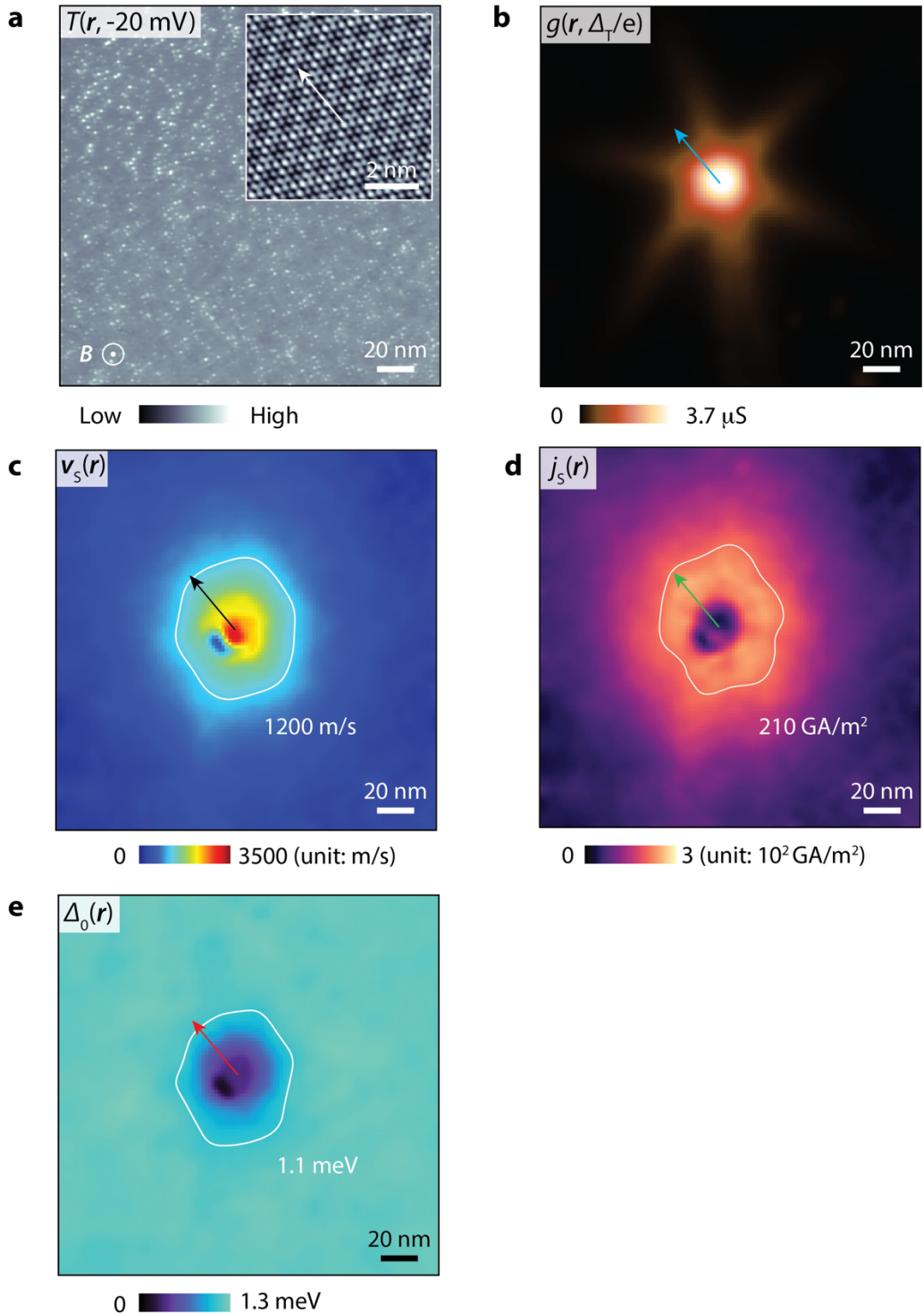

**Supplementary Figure 12.** Measured **a**, $T(r, -20\text{ mV})$, **b**, $g(r, \Delta_T/e)$, **c**, $v_S(r)$, **d**, $j_S(r)$, and **e**, $\Delta_0(r)$ in the same field of view. The contour lines in **c-e** show hexagonal symmetry in $v_S(r)$, $j_S(r)$, and $\Delta_0(r)$ with arrows in each image indicating their alignment with the crystal lattice (inset of **a**).